\RequirePackage{fixltx2e}
\documentclass[aps,prb,reprint,twocolumn,amsmath,
amssymb,amsfonts,superscriptaddress]{revtex4-1}

\usepackage{graphicx,xcolor}
\usepackage{dcolumn}
\usepackage{color}

\usepackage{hyperref}
\usepackage{microtype}
\usepackage{wasysym}
\usepackage{times}
\usepackage[varg]{txfonts}

\usepackage{mathrsfs}
\usepackage{xcolor}
\usepackage{graphicx}
\usepackage[percent]{overpic}
\usepackage{bm}
\usepackage{fixltx2e}
\usepackage{natbib}
\usepackage{braket}

\newcommand{\subref}[2]{\ref{#1}\hyperref[#1]{#2}}

\renewcommand{\vec}[1]{{\boldsymbol{#1}}}
\newcommand{\vhat}[1]{\vec{\hat{#1}}}

\newcommand{\tsub}[1]{${}_{#1}$}

\newcommand{\abo}[2]{#1\tsub{2}#2\tsub{2}O\tsub{7}}
\newcommand{\eto}{\abo{Er}{Ti}}
\newcommand{\hto}{\abo{Ho}{Ti}}
\newcommand{\yto}{\abo{Yb}{Ti}}
\newcommand{\tto}{\abo{Tb}{Ti}}
\newcommand{\dto}{\abo{Dy}{Ti}}

\newcommand{\hc}{{\rm h.c.}}
\newcommand{\avg}[1]{\langle #1 \rangle}
\newcommand{\im}{{\rm Im}}

\newcommand{\h}[1]{{#1}^{\dagger}} 
 
\newcommand{\cc}[1]{{#1}^{*}}

\newcommand{\sS}{\psi}
\newcommand{\sSA}{\psi A}
\newcommand{\sE}{E}

\newcommand{\exch}{{\rm ex}}

\newcommand{\QSI}{{\rm QSI}}
\newcommand{\VA}{V}
\newcommand{\VAn}[1]{V_{#1}}

\hypersetup{colorlinks=true,linkcolor=blue,citecolor=blue,urlcolor=blue} 

\begin{document}

\title{Fingerprints of quantum spin ice in Raman scattering}

\author{Jianlong Fu}
\affiliation{School of Physics and Astronomy, University of Minnesota, Minneapolis,
MN 55116, USA}

\author{Jeffrey G. Rau}
\affiliation{Department of Physics and Astronomy, University of Waterloo, Waterloo, ON, N2L 3G1, Canada}

\author{ Michel J. P. Gingras}
\affiliation{Department of Physics and Astronomy, University of Waterloo, Waterloo, ON, N2L 3G1, Canada}
\affiliation{Canadian Institute for Advanced Research, 180 Dundas St. W., Toronto, ON,  M5G 1Z8, Canada}.

\author{Natalia  B. Perkins}

\affiliation{School of Physics and Astronomy, University of Minnesota, Minneapolis,
MN 55116, USA}

\begin{abstract}
 We develop a theory of the dynamical response of a minimal model of
quantum spin ice (QSI) by means of inelastic light scattering.  In
particular, we are interested in the Raman response of the
fractionalized U(1) spin liquid realized in the XXZ QSI. We show
that the low-energy Raman intensity is dominated by spinon and gauge
fluctuations. We find that the Raman response in the QSI state of that
model appears only in the $T_{2g}$ polarization channel. We show that
the Raman intensity profile displays a broad continuum from the
spinons and coupled spinon and gauge fluctuations, and a low-energy
peak arising entirely from gauge fluctuations. Both features originate
from the exotic interaction between photon and the fractionalized
excitations of QSI. Our theoretical results suggest that inelastic
Raman scattering can in principle serve as a promising experimental
probe of the nature of a U(1) spin liquid in QSI.
\end{abstract}
\maketitle

Quantum spin liquids (QSLs) have proven to be one of the most
fascinating and challenging subjects in modern condensed matter
physics.\cite{Anderson1973, Wen2002,
Balents2010,Gingras2014,Savary2017} They are known to host a
remarkable set of emergent phenomena, including long-range
entanglement, topological ground state degeneracy and a number of
unusual fractionalized excitations such as fermionic or bosonic
spinons as well as emergent gauge excitations.  In recent years, there
has been significant progress both in the theoretical understanding of
such phenomena and in identifying realistic microscopic models that
may host QSL phases.  Notable examples include the spin-1/2 Heisenberg
antiferromagnet on the kagome
lattice,\cite{White2011,Depenbrock2012,Han2012,Iqbal2013,Ioannis2014}
the family of exactly solvable Kitaev-type
models,\cite{Kitaev2006,Mandal2009,Hermanns2015} and quantum spin
ice.\cite{Hermele2004,Molavin2007,Onoda2011,Benton2012,Ross2011,Savary2012,SBLee2012,
Gingras2014,Hao2014,Rau2015,Wang2016,Savary2016}
 
Direct experimental observation and characterization of QSLs is
challenging.  Unlike states with spontaneously broken symmetry, the
topological order characteristic of QSLs\cite{Savary2017} cannot be
captured by a local order parameter and thus cannot be directly
detected by local measurements. Identifying QSLs thus requires finding
experimental probes that provide information beyond the measurement of
local order parameters. One of the most promising avenues in this
direction is the characterization of the excitations of QSL
candidates. The fractionalized excitations of a QSL can be probed by
conventional methods such as inelastic neutron
scattering,\cite{Knolle2015} Raman
scattering\cite{Ko2010,Knolle2014,Brent2015,Brent2016-short,Brent2016-long}
or resonant X-ray scattering (RIXS),~\cite{
Ko2011,Gabor2016,Savary2015} all offering signatures that enable their
detection. Due to their fractionalized nature, these kinds of
scattering probes necessarily create multi-particle excitations in the
system. The appearance of such multi-particle continua in their
dynamical response is a hallmark of QSL behavior.
\cite{Wulferding2010,Sandilands2015,Sandilands2016,Yogesh2016,Glamazda2016,Gupta2016}
These continua are in stark contrast with the excitation spectra of
conventionally ordered phases, where sharp single-particle excitations
are expected.  Given the field currently still lacks the experimental
methods to probe the topological order of QSLs, it is therefore
important to have both a qualitative and a quantitative understanding
of these multi-particle continua and how they manifest themselves in
various experimental scattering probes and in QSL candidates.

\begin{figure}
\includegraphics[width=0.99\columnwidth] {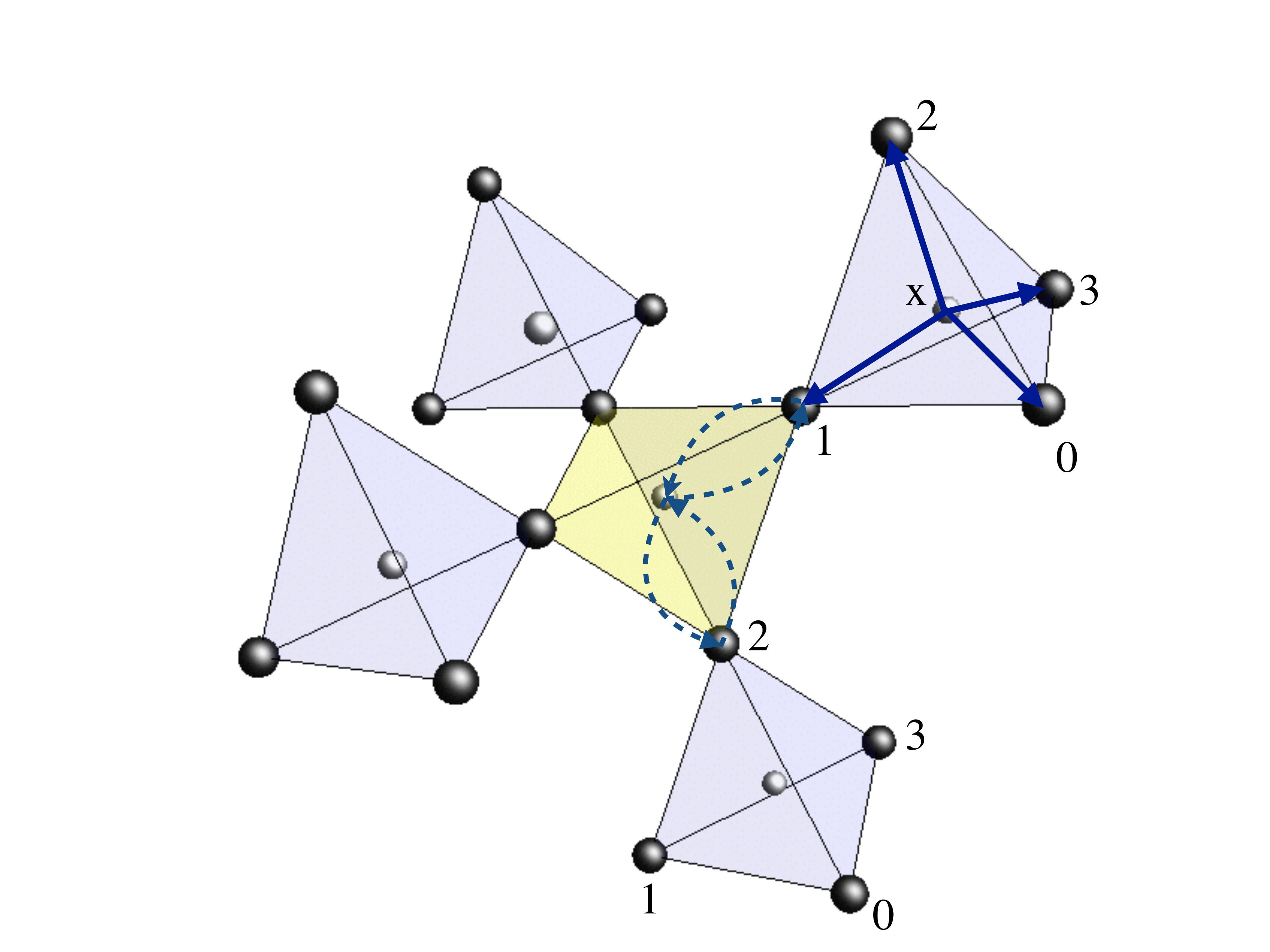}
\caption{
The pyrochlore lattice relevant for QSI materials. 
The centers of blue and yellow tetrahedra, labeled by
$\boldsymbol{x}$, form the $ \langle A \rangle $ and $ \langle B\rangle $ sublattices of the diamond
lattice, correspondingly.  $\mu=0,1,2,3$ label the bonds of the
diamond lattice. The spins, $S_{\boldsymbol{x},\mu}$, reside on
the pyrochlore sites located on the middle of the bond $\mu$.
The dashed lines illustrate the  electron hopping paths involved in the super-exchange
processes that generate the Raman vertex.
}
\label{spinicefigure}
\end{figure}

In this paper, we study such a dynamical response in a model QSL,
quantum spin ice (QSI). Defined on the pyrochlore lattice, a network
of corner-sharing tetrahedra (see Fig.\ref{spinicefigure}), this QSL
emerges naturally from the classical spin ice
limit.\cite{Balents2010,Gingras2014,bramwell2001spin} In this limit,
there are a macroscopic number of ground states characterized by the
so-called ``ice rule"; each tetrahedron must be in a two-in/two-out
state.\cite{bramwell2001spin} Excitations about this manifold have
three spins up and one down (or vice-versa) and can be separated at no
energy cost.\cite{Castelnovo2008} As first shown by
\citet{Hermele2004}, adding transverse exchange induces quantum
tunneling between different ice states. A sufficiently weak tunneling
stabilizes a QSL ground state with an emergent U(1) gauge field and
bosonic spinon excitations.\cite{Hermele2004,Savary2012,SBLee2012,
Gingras2014,Hao2014} Much effort has been put forth to understand the
nature of the QSI phase as well as its static and dynamic properties.
\cite{Hermele2004,banerjee2008,Ross2011,Savary2012,
SBLee2012,shannon2012quantum,Benton2012,Hao2014,kato2015numerical,
mcclarty2015chain,kwasigroch2016semi}

These theoretical studies have sparked intense experimental activity
aiming to find a concrete realization of QSI. The wide range of
rare-earth pyrochlore materials~\cite{Gardner2010,Gingras2014} have
provided an ample playground for this search. Potential candidates for
hosting a QSI phase currently include \tto{}, \yto{}, the
\abo{Pr}{$M$} family ($M$ = Zr, Sn, Hf) as well as the canonical
classical spin ices \dto{} and \hto{} (see
Ref.~[\onlinecite{Gingras2014}] for a survey). However, the physics of
these materials is complex; for many, it is even unclear how close
they are to the classical spin ice limit. Identifying experimental
probes that are sensitive to both the gauge and spinon excitations that
manifest in QSI would thus be useful for a better characterization of
these QSI candidates. Perhaps more importantly, it would deepen our
general understanding of the dynamical response of QSLs and their
various excitations.

In this article, we propose that inelastic Raman scattering may be of
particular interest for QSI systems. In a loose sense, we are inspired
by rather recent works on Raman scattering from Kitaev
QSLs.\cite{Ko2010,Knolle2014,Brent2015,Brent2016-short,Brent2016-long}
Using photons as a probe, the Raman response can in principle offer
insights in the excitation spectrum of a QSI that may not be
accessible through usual methods such as inelastic neutron scattering.  We
derive the Raman vertex for relevant rare-earth pyrochlore materials
using the traditional framework of an effective Hamiltonian for the
interaction of light with spin degrees of
freedom.\cite{LF1968,Shastry1990,Shastry1991,Ko2010} Applying these
results to an effective theory of QSI, we show how the gapped and
deconfined spinons as well as emergent gapless gauge modes appear in
the Raman spectrum.  Intriguingly, we find that real light can scatter
from the emergent ``light'' of QSI\cite{Benton2012} and produce a
measurable response. In addition, the spinon excitations themselves
have a direct signature in the Raman spectrum.

The structure of the paper is as follows: in Sections \ref{sec:model},
\ref{sec:qsi} and \ref{sec:spinon-gauge}, we set our notations and
review the basic concepts of QSI and its elementary
excitations.\cite{Hermele2004,Ross2011,Savary2012,Benton2012,Hao2014}
In Section \ref{sec:raman}, we briefly review the Loudon-Fleury theory
of the Raman scattering in Mott insulators that we need for our
study. Armed with this, we then derive the relevant
Raman operator involving the super-exchange processes between
pseudo-spins that represent the magnetic degrees of freedom. By
studying the polarization dependence of the Raman response, we
explicitly show that the response occurs only in the $T_{2g}$
polarization channel.  In Section \ref{sec:raman-qsi}, we compute the
Raman response for the XXZ QSI.  In particular, we first separate
three contributions to the Raman response -- from pure spinon
excitations, from gauge fluctuations and from their hybridization --
and then present the numerical results for the total Raman intensity
in the $T_{2g}$ channel.  Some discussion and a conclusion are given
in Secs. \ref{sec:discussion} and \ref{sec:conclusions}. 

\section{ Spin Hamiltonian}\label{sec:model}
Before delving into the details of the Raman process, we first review
the relevant anisotropic exchange models for the pertinent pyrochlore
materials.  In the current materials of interest which may realize
QSI, the magnetic degrees of freedom originate from rare-earth
ions.\cite{Gingras2014} Although we are not per se confining ourselves
to the details of the rare-earth ions that form the majority of the
QSI materials, it is useful to set the stage and make some general
observations about the spin Hamiltonian so far considered in the
theoretical and experimental investigations of QSI systems.

In the rare-earth ions,
the atomic interactions dominate; the free-ion ground state is
determined by following Hund's rules, first minimizing the Coulomb
energy, followed by the spin-orbit energy. These free-ion states have
well defined total angular momentum, $J$, and (approximately)
well-defined total orbital and spin angular momenta. In a crystalline
environment, due to the  electric fields from the
surrounding ions, the remaining $2J+1$ degeneracy of this manifold 
is partially lifted. When $J$ is a half-odd-integer, only Kramers' degeneracy
remains and one has a series of doublets for the relevant
$D_{3d}$ site symmetry~\cite{Gardner2010}. With respect to this
symmetry, these can transform either like spin-1/2 objects,
``pseudo-spin'' doublet (as in \yto{} or \eto{}), or like a more
exotic ``dipolar-octupolar'' doublet~\cite{huang2014quantum} (as in
\dto{} and Nd$_2$Zr$_2$O$_7$).  For integer $J$, Kramers' theorem does not apply and singlet
states are possible.  However, the $D_{3d}$ site symmetry can allow a
non-magnetic doublet, a so-called non-Kramers doublet (as, for
example in \hto{} or \tto{}). If well separated from the other crystal
field levels, each of these kinds of crystal field doublets behaves
like an effective spin-1/2 degree of freedom. For this reason we will
refer to all of these states as a ``spin'' regardless of whether they are
pseudo-spin-1/2, dipolar-octupolar or non-Kramers type.

To describe these doublets, we introduce the spin operators
$\vec{S}_i$, defined in the local basis at each site.\cite{Ross2011} For the
dipolar-octupolar and non-Kramers doublets, only $S^z_i$ contributes to the
magnetic dipole moment with $\vec{\mu}_i = -g \mu_B S^z_i \vhat{z}_i$,
where $\vhat{z}_i$ is the local $[111]$ direction. For the
pseudo-spin-1/2 case, both the $\vhat{z}_i$ component and the
components perpendicular to $\vhat{z}_i$ contribute to the dipole
moment. Since these three types of doublets transform quite differently under
the lattice symmetries,\cite{SBLee2012,huang2014quantum}  the allowed exchange
interactions are generally distinct. The most general nearest-neighbor
anisotropic exchange model on the pyrochlore lattice can be written
as:\cite{Ross2011,Savary2012}
\begin{align}
  \label{H_0}
     H_{\exch} &= \sum_{\avg{ij}}\left[
 J_{zz} S^z_i S^z_j -
J_{\pm}\left(S^+_i S^-_j+S^-_i S^+_j\right)+\right.  \\ &
\left.J_{\pm\pm} \left(\gamma_{ij} S^+_i S^+_j+\hc \right) \nonumber
 +J_{z\pm}\left( \zeta_{ij} \left[ S^z_i S^+_j+ S^+_i S^z_j \right]+ \hc \right)\right], 
 \end{align}
where the matrices $\zeta_{ij} = -\cc{\gamma}_{ij}$ and $\gamma_{ij}$ are
defined in Appendix \ref{app:defn}.  For the case of a pseudo-spin-1/2 doublet,
all of these couplings are allowed. For a non-Kramers doublet, one
has $J_{z\pm}=0$ whereas for a dipolar-octupolar doublet, the phases
are absent, i.e. $\gamma_{ij} = 1$ and
$\zeta_{ij}=1$.\cite{huang2014quantum}  Microscopically, these kinds
of short-range anisotropic interactions can be generated by various
super-exchange mechanisms.\cite{Onoda2011,Rau2015}  If $J_{zz}>0$ and
$J_{\pm\pm}=J_{\pm}=J_{z\pm}=0$, one recovers classical spin ice.
\footnote{We are focusing on a nearest-neighbor description and exclude long-range dipolar interactions.}
Introducing a finite $J_{\pm}$ or $J_{\pm\pm}$ with $J_{zz} \gg
J_{\pm}, J_{\pm\pm}\gg J_{z\pm}$ induces quantum tunneling between
the ice states~\cite{Hermele2004,Savary2012} and
stabilizes~\cite{shannon2012quantum,Benton2012} a QSI ground 
state. \footnote{Note that finite $J_{z\pm}$, for a pseudo-spin-1/2
doublet, produces an ordered ferromagnetic state drawn from the ice
manifold~\cite{Savary2012}.  To obtain QSI, one can thus only include $J_{z\pm}$ in
concert with $J_{\pm}$ or $J_{\pm\pm}$.  For a dipolar-octupolar
doublet, $J_{z\pm}$ is entirely innocuous and can be removed by a local
redefinition of the doublet states.} While in \dto{} and \hto{} one
expects $J_{\pm}$ and $J_{\pm\pm}$ to be negligible,~\cite{Rau2015} in
other materials such as \yto{}, \eto{} or \tto{},  experiments strongly  suggest these couplings are significant.\cite{Ross2011,savary2012order,
zhitomirsky2012quantum,robert2015spin} Since we are interested in the spin ice
limit, we will restrict ourselves  to cases where $J_{zz}$ is
dominant and is antiferromagnetic ($J_{zz}>0$).  In the remainder of the paper, we
thus work with the dimensionless  ratios
\begin{align} j_{\pm}&=J_{\pm}/J_{zz}, & j_{z\pm}&=J_{z\pm}/J_{zz}, &
j_{\pm\pm}&=J_{\pm\pm}/J_{zz},
\end{align}
which we assume to be small.

\section{Quantum spin ice}
\label{sec:qsi}
We now review the slave-particle description of QSI,~\cite{Savary2012}
using the formulation introduced in Ref.~[\onlinecite{Hao2014}].  In
the following, we use the notation of
Refs.~[\onlinecite{Savary2012},\onlinecite{Hao2014}] and label the
pyrochlore sites by a combined index $(\vec{x},\mu)$, in which
$\vec{x}$ denotes a diamond lattice site belonging to sublattice
$\langle A\rangle$ and $\mu=0,1,2,3$ are the four nearest neighbors of
the diamond site, as shown in Fig.~\ref{spinicefigure}. The spin at
the center of the bond $\avg{\vec{x},\vec{x} + \vec{\mu}}$ is then
labeled as $\vec{S}_{\vec{x},\mu}$, with $\vec{\mu}$ being the vector
connecting the two neighboring diamond sites shown in
Fig.~\ref{spinicefigure}.

The slave-particle formulation we will use is built on extending the
original Hilbert space to track the spin ice charge $Q_{\bf{x}}$ of
the diamond lattice sites independently of the spins of the pyrochlore
sites. In terms of the spins, the charges are defined as
\begin{eqnarray}
\label{Q}
Q_{\vec{x}}=\left\{\begin{array}{lll}
+\sum_{\mu}S_{\vec{x}, \mu}^{z} ,  & \vec{x} \in \avg{ A},   \\
-\sum_{\mu}S_{\vec{x}-\vhat{\mu}, \mu}^{z},  &\vec{x} \in \avg{B}.
\end{array} \right.
\end{eqnarray}
The charge operator $Q_{\vec{x}}$ characterizes violations of the ice
rules: $Q_{\vec{x}} = 0$ being satisfied for a two-in/two-out state,
while tetrahedra with three-in/one-out or three-out/one-in have
$Q_{\vec{x}} = \pm 1$ and those with all-in/all-out have $Q_{\vec{x}}
= \pm 2$. 
 
We study the exchange Hamiltonian, Eq.~(\ref{H_0}), in an enlarged Hilbert space
containing  both the charge and the spin degrees of freedom separately.  We
construct this by first introducing a new Hilbert space for the charge
operator $Q_{\vec{x}}$ that is now taken as distinct from the spins.
Second, we enlarge the range of allowed charges from strictly $0,\pm 1$ and $\pm 2$
to include all integers.  Explicitly, if we define the physical
Hilbert space as $\mathcal{H}_{\rm phys} = \bigotimes_{\vec{x},\mu}
\mathcal{H}_{1/2}$, where $\mathcal{H}_{1/2}$ is the spin Hilbert
space, then the extended space is
\begin{equation}
  \mathcal{H}_{\rm ext} = 
  \left[\bigotimes_{\vec{x},\mu} \mathcal{H}_{1/2}\right] \otimes
  \left[ \bigotimes_{\vec{x}} \mathcal{H}_{O(2)}\right]
   \equiv \mathcal{H}_s \otimes \mathcal{H}_Q ,
\end{equation}
and where $\mathcal{H}_{O(2)}$ is the Hilbert space of an $O(2)$ rotor,
defined at each diamond site and spanned by an infinite set of basis
states that satisfy $Q_{\vec{x}} \ket{q_{\vec{x}}} = q_{\vec{x}}
\ket{q_{\vec{x}}}$, where $q_{\vec{x}}$ is an integer. We define the
physical subspace as the one in which the $Q_{\vec{x}}$ operators satisfy
the constraint of Eq.~(\ref{Q}).

In this extended space, one then introduces a phase
$\theta_{\vec{x}}$, conjugate to the charge operators
$Q_{\vec{x}}$.~\cite{Savary2012} These two operators satisfy the canonical
commutation relations
\begin{equation}
  [\theta_{\vec{x}},Q_{\vec{x}'}] = i \delta_{\vec{x},\vec{x}'}.
\end{equation}
The quantization of $Q_{\boldsymbol{x}}$ implies the periodicity of
$\theta_{\boldsymbol{x}}$.  The operators $Q_{\boldsymbol{x}}$ and
$\theta_{\vec{x}}$ allow us to introduce a spinon operator,
$\psi_{\boldsymbol{x}}$, which is the basic element in a slave
particle description of spin ice.  To be precise, we define the
raising and lowering operators $\h{\psi}_{\vec{x}} =
e^{+i\theta_{\vec{x}}}$ and ${\psi}_{\vec{x}} =
e^{-i\theta_{\vec{x}}}$, satisfying
\begin{subequations}
  \begin{align}
    [\h{\psi}_{\vec{x}},Q_{\vec{x}'}] &= -\h{\psi}_{\vec{x}} \delta_{\vec{x},\vec{x}'},\\
    [\psi_{\vec{x}},Q_{\vec{x}'}] &= +\psi_{\vec{x}} \delta_{\vec{x},\vec{x}'}, 
  \end{align}  
\end{subequations}
which thus increase or decrease the charge quantum number at diamond
lattice site $\vec{x}$.  We then interpret $Q_{\vec{x}}$ as the spinon
number operator in the quantum theory, with $\h{\psi}_{\vec{x}}$ and
${\psi}_{\vec{x}}$ being spinon creation and annihilation
operators,~\cite{Savary2012} which live in the Hilbert space
$\mathcal{H}_Q$.

For the $\mathcal{H}_s$ part of the extended Hilbert space, we define
\emph{new} auxiliary spin-1/2 operators, $\vec{s}_{\vec{x},\mu}$. The
original physical spin-1/2 operators $\vec{S}_{\vec{x},\mu}$ can be
expressed in terms of the $\vec{s}_{\vec{x},\mu}$,
$\h{\psi}_{\vec{x}}$ and ${\psi}_{\vec{x}}$ operators as
\begin{subequations}
  \label{transformation}
  \begin{align}
    S_{\vec{x},\mu}^{+}&=\psi_{\vec{x}}^{\dagger}s_{\vec{x},\mu}^{+}\psi_{\vec{x}+\vhat{\mu}},\\
    S_{\vec{x},\mu}^{-}&=\psi_{\vec{x}+\vhat{\mu}}^{\dagger}s_{\vec{x},\mu}^{-}\psi_{\vec{x}},\\    S_{\vec{x},\mu}^{z}&=s_{\vec{x},\mu}^{z}.
  \end{align}
\end{subequations}
These combinations of operators are chosen such that the canonical
commutation relations of the original spin-1/2 operators,
$\vec{S}_{\vec{x},\mu}$, are preserved, and the physical constraint
defined by Eq.~(\ref{Q}) is also respected. If we were able to enforce
these constraints exactly, Eqs.~(\ref{Q}-\ref{transformation}) would
then constitute an exact reformulation of the original spin-1/2
problem of Eq.~(\ref{H_0}).  While such an exact description is not feasible, this set of variables have nevertheless proven\cite{Savary2012,Hao2014} to be a
useful starting point for describing the QSI phases of the anisotropic
exchange model given in Eq.~(\ref{H_0}).

The enlargement of the
Hilbert space implies a large degree of redundancy in this
description. In particular, note that the mapping defined by
Eq.~(\ref{transformation}) is invariant under the U(1) transformation
\begin{align}
\psi_{\vec{x}}&\rightarrow\psi_{\vec{x}}e^{i\alpha_{\vec{x}}},&
s^{\pm}_{\vec{x},\mu} \rightarrow s^{\pm}_{\vec{x},\mu} e^{\pm i(\alpha_{\vec{x}} - \alpha_{\vec{x}+\vhat{\mu}})}
\end{align}
for an arbitrary local phase factor $\alpha_{\vec{x}}$.
This gauge redundancy can be made explicit by recasting the
$\vec{s}_{\vec{x},\mu}$ operators in terms of an emergent gauge field,
${A}_{\vec{x},\mu}$, and an emergent electric field,
$E_{\vec{x},\mu}$, via
\begin{align}
  s^{\pm}_{\vec{x},\mu} &= |s^{\pm}_{\vec{x},\mu}| e^{\pm i A_{\vec{x},\mu}},
  & s^z_{\vec{x},\mu} &= E_{\vec{x},\mu}.
\end{align}
To simplify the problem, we replace the transverse components of the
spin operator by their average value, with $|s^{\pm}_{\vec{x},\mu}|
\approx \avg{ |s_{\vec{x},\mu}^{\pm}|}$, and only keep the phase of $
s^{\pm}_{\vec{x},\mu}$ as dynamical variable.~\cite{Wen2002} It is
easy to check that the electric field and the gauge field satisfy the
commutation relation
\begin{equation}
  [A_{\vec{x},\mu},E_{\vec{x}',\nu}]=i\delta_{\vec{x}\vec{x}'}\delta_{\mu\nu}.
\end{equation}
By construction, these fields are compact given the redundancy built
into the definition of $A_{\vec{x},\mu}$ and the periodicity of
$\theta_{\vec{x}}$.  This kind of mapping of an auxiliary spin-1/2
system to a gauge theory has been explored in many contexts; we refer
the reader to the literature for further
details~\cite{Hermele2004,Benton2012}.

Having performed this reformulation of the original spin degrees of freedom, we
now rewrite $H_{\exch}$ in terms of these new variables. One finds
\begin{widetext}
\begin{align}
\label{H_1}
H_{\exch} =&\frac{1}{2}\sum_{\vec{x}}Q_{\vec{x}}^{2} - j_{\pm}\avg{ s^{\pm}}^{2}\sum_{\vec{x}\in \avg{A}}\sum_{\mu<\nu}\big[\psi_{\vec{x}}^{\dagger}e^{i(A_{\vec{x},\mu}-A_{\vec{x}+\vhat{\mu}-\vhat{\nu},\nu})}\psi^{}_{\vec{x}+\vhat{\mu}-\vhat{\nu}}+\psi_{\vec{x}+\vhat{\mu}}^{\dagger}e^{-i(A_{\vec{x},\mu}-A_{\vec{x},\nu})}\psi^{}_{\vec{x}+\vhat{\nu}}+\hc\big] \nonumber \\
&-j_{z\pm}\avg{ s^{\pm}}\sum_{\vec{x}\in \avg{A}}\sum_{\mu\neq\nu}\big[E_{\vec{x},\mu}(\psi_{\vec{x}}^{\dagger}e^{iA_{\vec{x},\nu}}\psi^{}_{\vec{x}+\vhat{\nu}}+\psi_{\vec{x}+\vhat{\mu}-\vhat{\nu}}^{\dagger}e^{iA_{\vec{x}+\vhat{\mu}-\vhat{\nu},\nu}}\psi^{}_{\vec{x}+\vhat{\mu}})\zeta_{\mu\nu}+\hc\big]
\nonumber \\
           &-j_{\pm\pm}\avg{ s^{\pm}}^{2}\sum_{\vec{x}\in \avg{A}}\sum_{\mu<\nu}\big[(\psi_{\vec{x}}^{\dagger}\psi^{}_{\vec{x}+\vhat{\mu}}\psi_{\vec{x}}^{\dagger}\psi^{}_{\vec{x}+\vhat{\nu}}+\psi_{\vec{x}+\vhat{\mu}-\vhat{\nu}}^{\dagger}\psi^{}_{\vec{x}+\vhat{\mu}}\psi_{\vec{x}}^{\dagger}\psi^{}_{\vec{x}+\vhat{\mu}})\gamma_{\mu\nu}+\hc \big].
\end{align}
\end{widetext}
Here we see that the $J_{zz}$ and $J_{\pm}$ parts of $H_{\rm ex}$
describe the spinon degrees of freedom, as well as their interaction with the gauge
field $A$. Including finite $J_{z\pm}$ introduces further spinon-gauge
couplings, while $J_{\pm\pm}$ produces direct four-spinon
interactions.  In the current work we consider only
$J_{z\pm}=J_{\pm\pm}=0$. Focusing on this limit has several
advantages; aside from being theoretically simpler, this limit is shared
among the exchange models for all three types of microscopic degrees
of freedom discussed in Sec.~\ref{sec:model}.
\footnote{In addition, since there is no sign problem for the exchange
model when $J_{\pm} > 0$ and $J_{\pm\pm}=J_{z\pm}=0$. This would in
principle allow validation of these results though direct numerical
simulation.\cite{banerjee2008,kato2015numerical} However, given the difficulty and
opacity of such calculations, here we pursue a more analytical
route. }

As it stands, the reformulated model $H_{\rm ex}$ lacks any dynamics
for the gauge fields at leading order.  To remedy this, we follow
Ref.~[\onlinecite{Hao2014}] and add to the model 
\begin{equation}
  \label{eq:h-gauge}
  H_g \equiv \frac{U}{2} \sum_{\vec{x}\in
    \avg{A},\mu} E_{\vec{x},\mu}^{2}-g \sum_{\hexagon}\cos\left(\sum_{\vec{x}\mu \in\hexagon}A_{\vec{x},\mu}\right),  
\end{equation}
to endow the gauge sector with its own dynamics. We denote the
full model, with this additional gauge part as
\begin{equation}
  \label{eq:h-qsi}
  H_{\QSI} \equiv H_{\exch} + H_g.
\end{equation}
This final part is inspired from the form of the effective
Hamiltonian that arises when considering the effects of transverse
exchange on the ground state spin-ice manifold. The ``ring''-exchange
term, proportional to $g$ in Eq.~(\ref{eq:h-gauge}), appears first
at third order in $J_{\pm}$ or at sixth order in $J_{\pm\pm}$.~\cite{Hermele2004} This
effective model has been analyzed in detail in
Refs.~[\onlinecite{Hermele2004}, \onlinecite{Benton2012}]. Here we
have added it by hand to make up for some deficiencies in the
slave-particle approach. In terms of the $A_{\vec{x}\mu}$, this second
term describes the ``lattice curl" of the gauge field, while the first
term penalizes large electric fields, as required for the mapping of
the auxiliary spin-1/2 spins, $s_{\vec{x}\mu}$, to a gauge theory.  For our purposes, we
will assume the compactness of the gauge field is innocuous; namely
the effects of the gauge monopoles~\cite{Hermele2004,Kogut1979} are
not considered. Consistent with this assumption, we also take
$A_{\vec{x},\mu} \ll 1$. Under this condition, $H_{g}$ can be expanded
to give~\cite{Benton2012,Hao2014}
 \begin{equation}
 H_{g}=\sum_{\vec{x}\in\langle A\rangle,\mu}\left[
 \frac{U}{2}E_{\vec{x},\mu}^{2}+\frac{g}{2}B_{\vec{x},\mu}^{2}\right],
 \end{equation}
where the magnetic fluxes $B_{\vec{x},\mu}$ derive from the lattice
curl of the gauge field~\cite{Benton2012} $A_{\vec{x},\mu}$.  In such
a phenomenological description, the magnitudes of $U$ and $g$ must be
set by comparison with more precise calculations within the full
model.  For the case of $j_{\pm\pm}=j_{z\pm}=0$ they have been
estimated\cite{Benton2012,Hao2014} to be on the scale of $\sim
j_{\pm}^{3}$. More specifically, we use the values of
Ref.~[\onlinecite{Hao2014}], given as
\begin{align}
  \label{g-U}
g&\simeq 24 j_{\pm}^{3}, &
U&\simeq 2.16 j_{\pm}^{3}.
\end{align}
\section{Spinon dynamics and gauge fluctuations}
\label{sec:spinon-gauge}
We now consider the physics of $H_{\QSI}$ [Eq.~(\ref{eq:h-qsi})] in the XXZ limit, where
$j_{\pm\pm}=j_{z\pm}=0$. To simplify the spinon-gauge coupling, we
first expand in $A_{\vec{x}\mu}$, considering only the leading terms
in the expansion of $e^{iA_{\vec{x},\mu}}$. This will facilitate the
perturbative calculations to follow. Defining
$\tilde{j}_{\pm}=j_{\pm}\avg{ s^{\pm}}^{2}$, we can write
\begin{widetext}
\begin{eqnarray}
\begin{aligned}
\label{HXXZ}
H_{\QSI}&\sim \left[\frac{1}{2}\sum_{\vec{x}}Q_{\vec{x}}^{2}-\tilde{j}_{\pm}\sum_{\vec{x}\in \langle A\rangle}\sum_{\mu<\nu}(\psi_{\vec{x}}^{\dagger}\psi^{}_{\vec{x}+\vhat{\mu}-\vhat{\nu}}+\psi_{\vec{x}+\vhat{\mu}}^{\dagger}\psi^{}_{\vec{x}+\vhat{\nu}}+\hc)\right] \\
&-\bigg[\tilde{j}_{\pm}\sum_{\vec{x}\in \avg{A}}\sum_{\mu<\nu}i(\psi_{\vec{x}}^{\dagger}\psi^{}_{\vec{x}+\vhat{\mu}-\vhat{\nu}}-\psi^{}_{\vec{x}}\h{\psi}_{\vec{x}+\vhat{\mu}-\vhat{\nu}})(A_{\vec{x},\mu}-A_{\vec{x}+\vec{\mu}-\vec{\nu},\nu})+i(\psi_{\vec{x}+\vhat{\mu}}^{\dagger}\psi^{}_{\vec{x}+\vhat{\nu}}-\psi_{\vec{x}+\vhat{\nu}}^{\dagger}\psi^{}_{\vec{x}+\vhat{\mu}})(A_{\vec{x},\nu}-A_{\vec{x},\mu})\bigg]+H_g
\\ &\equiv H_{\sS}+H_{\sSA}+ H_g.
\end{aligned}
\end{eqnarray} 
\end{widetext}
We have broken this Hamiltonian into three parts, two of which are
new: $H_{\sS}$ which describes the kinetic energy of the bosonic
spinons $\psi_{\vec{x}}$ and their ``charging'' energy $\sim
Q_{\vec{x}}^2$, and $H_{\sSA}$ which describes a minimal coupling
between the spinons and the emergent U(1) gauge field.

The spinon part, $H_{\sS}$, defines a quantum rotor model and is thus
difficult to solve even on its own. This can be written as
\begin{align}
  H_{\sS} &=
            \frac{1}{2}\sum_{\vec{x}}Q_{\vec{x}}^{2}-\tilde{j}_{\pm}\sum_{\vec{x}\in \avg{A}}\sum_{\mu<\nu}(\psi_{\vec{x}}^{\dagger}\psi^{}_{\vec{x}+\vhat{\mu}-\vhat{\nu}}+\psi_{\vec{x}+\vhat{\mu}}^{\dagger}\psi^{}_{\vec{x}+\vhat{\nu}}+\hc), \nonumber\\
          &= \label{eq:raman-s}
            \frac{1}{2}\sum_{\vec{x}}Q_{\vec{x}}^{2}
            -\tilde{j}_{\pm} \sum_{\mu<\nu} \sum_{\vec{k}\lambda} f^{\sS}_{\mu\nu}(\vec{k}) \h{\psi}_{\vec{k}\lambda} \psi^{}_{\vec{k}\lambda},
\end{align}
where we have introduced the sublattice label $\lambda=\avg{A},\avg{B}$ and defined the vertex
\begin{equation}
  \label{eq:form-s}
  f^{\sS}_{\mu\nu}(\vec{k}) \equiv 2 \cos{\left[\vec{k}\cdot\left(\vec{\mu}-\vec{\nu}\right)\right]}.
\end{equation}
To approximately solve
this rotor model, we use the ``exclusive boson" representation
introduced in Ref.~[\onlinecite{Hao2014}]
\begin{subequations}
\begin{align}
 \psi_{\vec{x}}&=\frac{d_{\vec{x}}+\h{b}_{\vec{x}}}{(1+\h{d}_{\vec{x}}d_{\vec{x}}+\h{b}_{\vec{x}}b_{\vec{x}})^{1/2}}, \\
Q_{\vec{x}}&=\h{d}_{\vec{x}}d_{\vec{x}}-\h{b}_{\vec{x}}b_{\vec{x}}.
\end{align}
\end{subequations}
Here $b_{\vec{x}}$ and $d_{\vec{x}}$ are bosonic operators constrained to satisfy $b_{\vec{x}}d_{\vec{x}}\equiv b_{\vec{x}}^{\dagger}d_{\vec{x}}^{\dagger}\equiv 0$ for all the basis states.  Under the approximation that the density of bosons is small, and thus  dropping all four-boson terms, the Hamiltonian $H_{\psi}$ is simplifies significantly into a quadratic form.
This can then be diagonalized with the help of a Bogoliubov transformation, giving
\begin{equation}
  H_{\psi}=\sum_{\vec{k}\lambda}
    E_{\vec{k}}
    \left(\h{\tilde{d}}_{\vec{k}\lambda}\tilde{d}^{}_{\vec{k}\lambda}+
    \h{\tilde{b}}_{\vec{k}\lambda}\tilde{b}^{}_{\vec{k}\lambda}\right) +{\rm const.}
\end{equation}
where $\tilde{b}_{\vec{k}\lambda}$, $\tilde{d}_{\vec{k}\lambda}$ are the Bogoliubov
quasi-particles and the dispersion relation
$E_{\vec{k}}$ is given by
  \begin{equation}    
\label{spinondispersion}
E_{\vec{k}}=\frac{1}{2}\left[
  1-2\tilde{j}_{\pm}\sum_{\alpha \neq \beta}\cos\left(
    \frac{k_{\alpha}}{2}\right)\cos\left(\frac{k_{\beta}}{2}\right)
\right]^{1/2},
\end{equation}
where $\alpha,\beta=x,y,z$
are the three global cubic directions.
Explicit expressions for the relationship between the spinons $\psi_{\vec{k}\lambda}$
and the bosons $\tilde{b}_{\vec{k}\lambda}$, $\tilde{d}_{\vec{k}\lambda}$ are given in Ref.~[\onlinecite{Hao2014}].
The Green's function for the spinon field~\cite{Hao2014} is then given by
\begin{align}
\label{eq:green-s}
G_{\sS}(\omega,\vec{k})&=\int dt\ e^{i\omega t} \left[-i\avg{ \mathcal{T} \psi^{}_{\vec{k}}(t)\psi_{\vec{k}'}^{\dagger}(0)} \right],\nonumber\\
                     &= \frac{1}{2E_{\vec{k}}}\left[
                       \frac{1}{\omega-E_{\vec{k}} + i\delta}-
                       \frac{1}{\omega+E_{\vec{k}} - i\delta}
                       \right],
%  &\equiv  \frac{1}{\omega^2-(E_{\vec{k}} - i\delta)^2}
\end{align} 
where ${\mathcal T}$ implements time-ordering and $\delta = 0^+$.

Next, we discuss the dynamics of the gauge Hamiltonian,
$H_{g}$. This can be done using standard
methods~\cite{Hao2014} once the condition $A_{\vec{x},\mu}\ll 1$ has
been imposed. Explicitly, one has
\begin{align}
  H_g  &\sim
  \frac{U}{2} \sum_{\vec{x}\in
    \avg{A},\mu} E_{\vec{x},\mu}^{2}+\frac{g}{2} \sum_{\hexagon}\left(\sum_{\vec{x}\mu \in\hexagon}A_{\vec{x},\mu}\right)^2.     
\end{align}
To diagonalize $H_{g}$, a linear
transformation is defined as
\begin{align}
A_{\vec{p},\mu}&=\sum_{\gamma=0,1}\eta_{\mu \gamma}(\vec{p})a_{\gamma,\vec{p}},
& E_{\vec{p},\mu}&=\sum_{\gamma=0,1}\eta_{\mu \gamma}(\vec{p})e_{\gamma,\vec{p}}, 
\end{align}
where $\eta_{\mu \gamma}(\vec{p})$
is a  matrix satisfying
\begin{equation}
\label{propertiesofeta}
\begin{aligned}
&\eta_{\mu \gamma}(-\vec{p})=\eta_{\mu \gamma}^{*}(\vec{p}), \\
& \sum_{\gamma}\eta_{\mu \gamma}(\vec{p})\eta_{\nu \gamma}(-\vec{p})=\sum_{\gamma}\eta_{\mu \gamma}(\vec{p})\eta_{\nu \gamma}^{*}(\vec{p})=
\delta_{\mu\nu}.
\end{aligned}
\end{equation}
The two operators, $\hat{a}_{\gamma,\vec{p}}$ and $ \hat{e}_{\gamma',\vec{p}}$,
satisfy the canonical commutation relation $\left[
\hat{a}_{\gamma,\vec{p}}, \hat{e}_{\gamma',\vec{p}'}\right]=i
\delta_{\vec{p},\vec{p}'}\delta_{\gamma,\gamma'}$. This way, the $a$-excitations act
like positions and $e$-excitations act like momenta in a quantum
harmonic oscillator. This unitary transformation  diagonalizes
$H_{g}$, resulting in
\begin{align} 
  H_{g}&=\sum_{\vec{p} \gamma}\left[ \frac{U}{2}\hat{e}_{\gamma,\vec{p}}\hat{e}_{\gamma,-\vec{p}}+
         \frac{\epsilon_{\vec{p}}^{2}}{2 U}\hat{a}_{\gamma,\vec{p}}\hat{a}_{\gamma,-\vec{p}}
         \right],
\end{align}
where we see that $ \hat{a}_{\gamma,\vec{p}}$ and $\hat{e}_{\gamma,\vec{p}}$ are
transverse modes ($\gamma=0,1$) describing the gauge fluctuations and dynamics of
electric fluxes, respectively.
The photon dispersion is defined as
\begin{align}
 \epsilon^2_{\vec{p}}&=4 Ug \left[ 3-\frac{1}{2} \sum_{\alpha\neq \beta} 
                       \cos\left(\frac{p_{\alpha}}{2}\right)\cos\left(\frac{p_{\beta}}{2}\right) \right] \\
  &\simeq c^2|\vec{p}|^2+O(|\vec{p}|^4), \nonumber
\end{align}
where $c = (Ug)^{1/2}$.  This speed of emergent light, $c\simeq 0.3
g$, has been estimated in simulations of the effective ring-exchange
model~\cite{Benton2012} and motivated the value of $U$ given in
Eq.~(\ref{g-U}).  The Green's functions for these $a$- and $e$-operators
can also be easily worked out.  One arrives at\cite{Hao2014}
\begin{subequations} 
\begin{align}
\label{eq:green-a}
  G_{A}(\omega,\vec{p})
  &=\int dt e^{i\omega t}
   \left[-i \avg{ \mathcal{T}\hat{a}_{\gamma,\vec{p}}(t)\hat{a}_{\gamma,-\vec{p}}(0)}\right],\nonumber\\
  &=\frac{U}{\omega^{2}-\epsilon_{\vec{p}}^{2}+i\delta}, \\
\label{eq:green-e}
  G_{E}(\omega,\vec{p})
  &=\int dt e^{i\omega t}\left[-i \avg{\mathcal{T}\hat{e}_{\gamma,\vec{p}}(t)\hat{e}_{\gamma,-\vec{p}}(0)}\right],\nonumber\\
  &=\frac{\epsilon_{\vec{p}}^{2}}{U(\omega^{2}-\epsilon_{\vec{p}}^{2}+i\delta)},
\end{align}
\end{subequations}
where $\delta = 0^+$.

Finally, we have the interaction between the spinons and 
gauge field encapsulated in $H_{\sSA}$.  This
interaction can be re-written in momentum space as
\begin{widetext}
\begin{eqnarray}
\label{vertexA}
\begin{aligned}
  H_{\sSA}&=-\tilde{j}_{\pm}\sum_{\mu<\nu}\sum_{\vec{x}\in \avg{A}}i\left[(\h{\psi}_{\vec{x}}\psi^{}_{\vec{x}+\vhat{\mu}-\vhat{\nu}}-\psi^{}_{\vec{x}}\h{\psi}_{\vec{x}+\vhat{\mu}-\vhat{\nu}})(A^{}_{\vec{x},\mu}-A^{}_{\vec{x}+\vec{\mu}-\vec{\nu},\nu})  +(\h{\psi}_{\vec{x}+\vhat{\mu}}\psi^{}_{\vec{x}+\vhat{\nu}}-\h{\psi}_{\vec{x}+\vhat{\nu}} \psi^{}_{\vec{x}+\vhat{\mu}})(A^{}_{\vec{x},\nu}-A^{}_{\vec{x},\mu})\right],\\  
  &\equiv -\frac{\tilde{j}_{\pm}}{\sqrt{N}}\sum_{\mu<\nu} \sum_{\vec{k}\lambda}\sum_{\rho}
  f^{\sSA}_{\mu\nu\rho,\lambda}(\vec{k},\vec{p}) \label{eq:raman-sa}
                             \h{\psi}_{\vec{k}+\vec{p},\lambda}\psi^{}_{\vec{k}\lambda}   A^{}_{\vec{p},\rho},\\
\end{aligned}
\end{eqnarray}
where $N$ is the number of unit cells 
and we have defined the vertex
\begin{align}
  f^{\sSA}_{\mu\nu\rho,\lambda}(\vec{k},\vec{p})
  =i\delta_{\lambda,\avg{A}}&\left[ 
     \delta_{\rho\mu} e^{+i \vec{k}\cdot(\vec{\mu}-\vec{\nu})}-
    \delta_{\rho\nu}e^{+i (\vec{k}+\vec{p})\cdot(\vec{\mu}-\vec{\nu})}
    +\delta_{\rho\nu} e^{-i \vec{k}\cdot(\vec{\mu}-\vec{\nu})}
    -\delta_{\rho\mu}e^{-i (\vec{k}+\vec{p})\cdot(\vec{\mu}-\vec{\nu})} \nonumber
      \right] +\\
  i\delta_{\lambda,\avg{B}}&\left[
     \delta_{\rho\mu} e^{+i (\vec{k}\cdot(\vec{\mu}-\vec{\nu}) - \vec{p}\cdot\vec{\nu})}
    -\delta_{\rho\nu}e^{+i (\vec{k} \cdot(\vec{\mu}-\vec{\nu}) - \vec{p}\cdot \vec{\nu})}
    +\delta_{\rho\nu} e^{-i (\vec{k}\cdot(\vec{\mu}-\vec{\nu}) +\vec{p}\cdot\vec{\mu})}
    -\delta_{\rho\mu}e^{-i (\vec{k}\cdot(\vec{\mu}-\vec{\nu})+\vec{p}\cdot\vec{\mu})}
    \right].    \label{eq:form-sa}
\end{align}
\end{widetext}
This part, $H_{\sSA}$, describes an interaction between the spinons
$\psi$ and the gauge field $A$, similar to the interaction in regular
quantum electrodynamics, coupling $A$ to the ``current'' of the
spinons. At this point, we thus have a theory of spinons interacting
with a U(1) gauge field.

\section{Microscopic origin of the Raman Vertex}
\label{sec:raman}
We now investigate the mechanism of light scattering from the
excitations of a QSI phase.  Light can interact with matter in various
ways.  It is well known that, in general, the strongest coupling does
not come from the direct coupling of the magnetic field of the light
with the magnetic moments, but rather through the coupling of its electric
field to the electric dipole moments of the scattering
medium.~\cite{LF1968,Hayes2012} The basic processes leading to the Raman
response in Mott insulators are similar to those leading to exchange
interactions, except that the virtual electron hopping is assisted by
photons. Consequently, in the simplest approximation, the operator
describing Raman processes is generically expected to be proportional
to the spin-exchange couplings, weighted by polarization-dependent
factors that determine the ability of the photons to control the
magnitude of an electron hopping along certain
bonds.~\cite{LF1968,Shastry1990,Shastry1991,Devereaux2007,Perkins2008,Perkins2013,Ko2010,Knolle2014}

To describe the coupling of light to electrons on a lattice,
one can, in a first approximation, perform a Peierls substitution,~\cite{hofstadter1976}
attaching a ``Wilson line" operator to the electron hopping term to preserve gauge
invariance~\cite{Shastry1990,Devereaux2007,Ko2010}
 \begin{equation}
\label{Wilson}
c_{i\sigma}^{\dagger}c^{}_{j\sigma}\rightarrow c_{i\sigma}^{\dagger}c^{}_{j\sigma}\exp \left[\frac{ie}{\hbar c}\int_{\vec{r}_j}^{\vec{r}_i} d\vec{r} \cdot \vec{\mathcal{A}}(\vec{r})\right]. 
\end{equation}   
Here we use $\vec{\mathcal{A}}$ to denote the vector potential of the
radiation field, not to be confused with the emergent U(1) gauge field
in QSI, which we have denoted as $A$. Intuitively, the photon couples
to the electric dipole formed by charge transfer between different
lattice sites.  Thus, in order to get the correct Raman vertex, we
must know the microscopic electron hopping mechanism at play in the
material.

In the case of QSI, the super-exchange interactions between
neighboring spins are expected to be mediated by the oxygen atoms that
surround each rare-earth ion\cite{Onoda2011,Rau2015} as illustrated in
Fig.~\ref{spinicefigure}. The microscopic derivation of
Eq.~(\ref{H_0}) starts from separating the total microscopic
Hamiltonian into an on-site part, $H_0$, and the hopping between
rare-earth $f$ electrons and oxygen $p$ electrons, $V_0$
\begin{equation}
{H}_{}={H}_0+{V}_0.
\end{equation}
All other hoppings are assumed to be small and thus
neglected.~\cite{Rau2015} The ${V}_{0}$ term is given by
\begin{align}
\label{V0}
  {V}_{0}=
 \sum_{\vec{x}\in\avg{A}}\sum_{\mu}\sum_{\alpha\beta}
  &\left(t_{\mu,\alpha\beta}^{\dagger}p_{\vec{x},\alpha}^{\dagger}f^{}_{\vec{x}\mu,\beta}+
    t^{\dagger}_{\mu,\alpha\beta}p_{\vec{x}+\vec{\mu},\alpha}^{\dagger}f^{}_{\vec{x}\mu,\beta}\right.
  \\+&\left.
     t_{\mu,\alpha\beta}f_{\vec{x}\mu,\beta}^{\dagger}p^{}_{\vec{x},\alpha}+
     t_{\mu,\alpha\beta}f_{\vec{x}\mu,\beta}^{\dagger}p^{}_{\vec{x}+\vec{\mu},\alpha}\right),\nonumber
\end{align}
where $t_{\mu,\alpha\beta}$ denotes the hopping amplitude,
$f_{\vec{x}\mu,\beta}$ and $p_{\vec{x},\alpha}$ represent the electron
annihilation operators on the rare-earth and oxygen ions, respectively.
Here we only include the high-symmetry oxygens, those which lie in the centers
of the rare-earth tetrahedra, as they are closer to the rare-earth ions than the
low symmetry oxygens~\cite{Gardner2010}. The on-site part $H_0$ contains the atomic
interactions of the rare-earth ion, including Coulomb, spin-orbit and crystal field
contributions. We do not need the detailed properties of $H_0$, save for that its ground
state is a doublet, as discussed in Sec.~\ref{sec:model}, and that the energy to add or remove an
electron, denoted roughly as $\sim U_f$, is large relative to the hoppings, $t$.

We now include the interaction with the electromagnetic (EM) field.
As mentioned in Eq.~(\ref{Wilson}), this coupling brings about a
modification, ${V}_{0}\rightarrow \VA$, given by
\begin{align}
\label{VA11}  \VA=\sum_{\vec{x}\in\avg{A}}\sum_{\mu}\sum_{\alpha\beta}&\left(t_{\mu,\alpha\beta}^{\dagger}p_{\vec{x},\alpha}^{\dagger}f^{}_{\vec{x}\mu,\beta}e^{\frac{ie}{\hbar c}\int_{\vec{x}+\frac{1}{2}\vec{\mu}}^{\vec{x}} d\vec{r}\cdot\vec{\mathcal{A}}(\vec{r})}\right.\nonumber\\
  +&\left.t^{\dagger}_{\mu,\alpha\beta}p_{\vec{x}+\vec{\mu},\alpha}^{\dagger}f^{}_{\vec{x}\mu,\beta}e^{\frac{ie}{\hbar c}\int_{\vec{x}+\frac{1}{2}\vec{\mu}}^{\vec{x}+\vec{\mu}} d\vec{r}\cdot \vec{\mathcal{A}}(\vec{r})}+\hc\right).
\end{align}
To proceed, we make the assumption that the photon
field is relatively weak, so that interaction with light does not
affect the electronic structure of the material.  We also assume that
$\frac{ie}{\hbar c}\vec{\mathcal{A}}\cdot \vec{\mu}$ is reasonably
small so that we can expand $\VA$ using a Taylor expansion as
\begin{equation}
\VA=\VAn{0}+\VAn{1}+\cdots.
\end{equation}
Knowing that the wavelength of the incoming and outgoing EM waves are much larger than the lattice constant of the material, we can further make the replacement
\begin{equation}
  \frac{ie}{\hbar c}\int_{\vec{x}+\frac{1}{2}\vec{\mu}}^{\vec{x}} d\vec{r}\cdot\vec{\mathcal{A}}(\vec{r}) \sim
  -\frac{ie}{2\hbar c}\left(\vec{\mu} \cdot \vec{\mathcal{A}}_{\vec{x}}\right).
\end{equation}
Under these approximations we have that
\begin{align}
\label{VA12}
  \VAn{1}=\left(\frac{ie}{2\hbar c}\right)
  \sum_{\vec{x}\in\avg{A}}\sum_{\mu,\alpha\beta}
  (\vec{\mathcal{A}}_{\vec{x}}\cdot\vec{\mu})
  \bigg[ &t_{\mu,\alpha\beta}^{\dagger}p_{\vec{x}+\vec{\mu},\alpha}^{\dagger}f^{}_{\vec{x}\mu,\beta}+
           t_{\mu,\alpha\beta}f_{\vec{x}\mu,\beta}^{\dagger}p^{}_{\vec{x},\alpha} \nonumber \\ \nonumber
  -&t_{\mu,\alpha\beta}^{\dagger}p_{\vec{x},\alpha}^{\dagger}f^{}_{\vec{x}\mu,\beta}
     -t_{\mu,\alpha\beta}f_{\vec{x}\mu,\beta}^{\dagger}p^{}_{\vec{x}+\vec{\mu},\alpha}\bigg].
\end{align}
This differs from $\VAn{0}$ in that it attaches to each electron hopping term a  factor $\pm \vec{\mathcal{A}}\cdot\vec{\mu}$ coming from the EM field.
In addition to this modification of the electron hopping, we also must
now include the energy of the EM field itself, which we denote as $H_{\gamma}$. 

Our goal is to derive an effective Hamiltonian, treating $V$ as a
perturbation, for the low-energy states of $H_0+H_{\gamma}$. For our
purposes this low-energy subspace contains all of the relevant EM
states and only the ground states of $H_0$. Now, from standard
degenerate perturbation theory,~\cite{lindgren1974rayleigh} this
effective Hamiltonian can be written
\begin{eqnarray}
\label{perturbationexpansion}
\begin{aligned}
{H}_{\rm eff}=&P{H}_{0}P +P{H}_{\gamma}P+{P}{V}{P}+{P}{V}{R}{V}{P}
\\&+
{P}{V}{R}{V}{R}{V}{P}
+{P}{V}{R}{V}{R}{V}{R}{V}{P}+\cdots,
\end{aligned}
\end{eqnarray}
in which ${P}$ projects into the ground state manifold of $H_0$ and
$R$ is the resolvent
\begin{equation}
{R}=\frac{1-{P}}{E_0-H_{0}-H_{\gamma}+i\delta}\approx \frac{1-{P}}{E_0-H_{0}+i\delta}  
\end{equation}
where $E_0$ is the ground state energy of $H_0$ and $\delta =
0^+$. Here, we have made the approximation that the energy of the
light, encoded in $H_{\gamma}$, is insignificant relative to the atomic
energy scales of $H_0$. We return to this effects of this
approximation in Sec.~\ref{sec:disc:micro}.  The presence of the
projection operator, ${P}$, implies that only even order perturbations
have non-zero contribution.  To get the non-resonant Raman vertex, we
neglect higher order terms and keep only $\VAn{1}$ as perturbation.

We now proceed to compute the effective Hamiltonian in the low-energy
subspace relevant for the rare-earth ion, similar to what is done in
calculations of super-exchange.\cite{Onoda2011,Rau2015} Due to the
structure of the super-exchange processes, the anisotropic exchange
Hamiltonian shown in Eq.~(\ref{H_0}) appears at fourth
order\cite{Onoda2011,Rau2015} in $\VAn{0}$, with
$H_{\exch}={P}\VAn{0}{R}\VAn{0}{R}\VAn{0}{R}\VAn{0}{P}$. It can be
shown that the Raman interaction also comes in at fourth order in
perturbation theory. To describe the scattering of light, we keep only the
leading $O(\mathcal{A}^2)$ parts of $H_{\rm eff}$, that is those
having two factors of $V_0$ and two factors of $V_1$.~\footnote{Note
that the higher order correction, $V_2$ would produce at the
same order, i.e. at $O(\mathcal{A}^2)$, from the first order
perturbative contribution $PV_2 P$. However, since $V_2$ still
necessarily involves a charge transfer from the ligand to the
rare-earth ion one has $PV_2P = 0$} While second-order processes,
$H^{(2)}_R$, that can contribute single-spin operators to the Raman
operator do exist (see Appendix \ref{app:single-ion}), they vanish
when only the high-symmetry oxygens are considered. We will return to
this point this is Sec.~\ref{sec:disc:micro}. To separate out the
Raman part, we then can write
\begin{widetext}
\begin{eqnarray}
\begin{aligned}
  {H}^{(4)}_R &\equiv {P}{V}{R}{V}{R}{V}{R}{V}{P}-H_{\exch},\nonumber \\
  &={P}(\VAn{0}+\VAn{1}){R}(\VAn{0}+\VAn{1}){R}(\VAn{0}+\VAn{1}){R}(\VAn{0}+\VAn{1}){P}-H_{\exch},\\
&={P}\VAn{1}{R}\VAn{1}{R}\VAn{0}{R}\VAn{0}{P}
+{P}\VAn{1}{R}\VAn{0}{R}\VAn{1}{R}\VAn{0}{P}+\cdots+{P}\VAn{0}{R}\VAn{0}{R}\VAn{1}{R}\VAn{1}{P}.
\end{aligned}\label{H4}
\end{eqnarray}
\end{widetext}
There are six terms that give a Raman contribution at fourth order in
$V$, with the two $\VAn{1}$ terms corresponding one incoming photon
and one outgoing photon. Analyzing these terms carefully, we find that
the Raman couplings can be derived by attaching photon factors, $\sim \pm
\mathcal{A} \cdot \vec{\mu}$, to two of the four hoppings in each
super-exchange process.

In light of this, we can write down ${H}^{(4)}_R$ explicitly.  We
first write $\vec{\mathcal{A}}_{\vec{x}}$ in terms of photon
operators, splitting it into two parts\cite{Shastry1990,Ko2010}
\begin{align}
 \vec{\mathcal{A}}_{\vec{x}}&\sim g_i \vhat{e}_{i}
                              {a}^{}_{\vec{k}_{i},\vhat{e}_{i}}e^{i\vec{k}_{i}\cdot \vec{x}}+g_f\vhat{e}_{f} {a}_{\vec{k}_{f},\vhat{e}_{f}}^\dagger e^{i\vec{k}_{f}\cdot \vec{x}}
  \equiv \vec{\mathcal{A}}_i + \vec{\mathcal{A}}_f.
\end{align}
 Here, ${a}^{}_{\vec{k}_{i},\vhat{e}_{i}}$ and $
{a}_{\vec{k}_{f},\vhat{e}_{f}}^\dagger$ represent the real photons,
not to be confused with the emergent excitations that exist in QSI.  The
vectors $\vhat{e}_{i}$ and $\vhat{e}_{f}$ denote the polarization
vectors of the incoming and outgoing photons, respectively.  The
 $g_i$ and $g_f$ are constant prefactors depending on the
incoming/outgoing photon frequencies. We will omit them in what follows.
Further, since the photon wave-vector is small relative to the inverse
lattice spacing, we can safely replace $ e^{i\vec{k}_{i}\cdot
\vec{x}}\sim e^{i\vec{k}_{f}\cdot \vec{x}}\sim 1$, keeping only
the polarization vectors $\vhat{e}_{i}$ and $\vhat{e}_{f}$.

For a hopping process involving bonds $\vec{\mu}$ and $\vec{\nu}$, the incoming and outgoing EM operator $\vec{\mathcal{A}}_{i}$ and $\vec{\mathcal{A}}_{f}$ can couple to $\pm\vec{\mu}$ and $\pm\vec{\nu}$. There are 12 possibilities in total for choosing two out of four bonds to couple with $\vec{\mathcal{A}}_{i}$ and $\vec{\mathcal{A}}_{f}$ (see Fig.~\ref{spinicefigure}). The overall prefactor is then found to be
\begin{eqnarray}
\begin{aligned}
&(\vec{\mathcal{A}}_{i}\cdot(\pm\vec{\mu}))(\vec{\mathcal{A}}_{f}\cdot(\mp\vec{\mu}))+(\vec{\mathcal{A}}_{i}\cdot(\pm\vec{\mu}))(\vec{\mathcal{A}}_{f}\cdot(\pm\vec{\nu}))\\&+(\vec{\mathcal{A}}_{i}\cdot(\pm\vec{\mu}))(\vec{\mathcal{A}}_{f}\cdot(\mp\vec{\nu}))+(\mu\rightleftharpoons\nu),\\
&\sim(\vec{\mathcal{A}}_{i}\cdot\vec{\mu})(\vec{\mathcal{A}}_{f}\cdot\vec{\mu})+(\vec{\mathcal{A}}_{i}\cdot\vec{\nu})(\vec{\mathcal{A}}_{f}\cdot\vec{\nu}).\nonumber
\end{aligned}
\end{eqnarray}
Since the Raman vertex and the effective Hamiltonian have similar mathematical form, we can easily express the Raman vertex in terms of  spin operators.\cite{Shastry1990,Ko2010} 
The final result for the Raman part of the effective Hamiltonian is given by
\begin{equation}
\label{ramanvertex}
H_R^{(4)} \sim \sum_{\mu<\nu}[(\vec{\mathcal{A}}_{i}\cdot\vec{\mu})(\vec{\mathcal{A}}_{f}\cdot\vec{\mu})+(\vec{\mathcal{A}}_{i}\cdot\vec{\nu})(\vec{\mathcal{A}}_{f}\cdot\vec{\nu})]{\mathcal{R}}_{\mu\nu},
\end{equation}
where the operator ${\mathcal{R}}_{\mu\nu}$ is defined as
\begin{widetext}
\begin{eqnarray}
\label{Roperator}
\begin{aligned}
{\mathcal{R}}_{\mu\nu}=\sum_{\vec{x} \in \avg{A}}&\left[J_{zz}\left(S_{\vec{x}\mu}^{z}S_{\vec{x}\nu}^{z}+S_{\vec{x}\mu}^{z}S_{\vec{x}+\vec{\mu}-\vec{\nu},\nu}^{z}\right)-
  J_{\pm}\left(S_{\vec{x}\mu}^{+}S_{\vec{x}\nu}^{-}+S_{\vec{x}\mu}^{+}S_{\vec{x}+\vec{\mu}-\vec{\nu},\nu}^{-}+\hc\right)-
\right.\\
&\left. J_{z\pm}\left(S_{\vec{x}\mu}^{z}(S_{\vec{x}\nu}^{+}+S_{\vec{x}+\vec{\mu}-\vec{\nu},\nu}^{+})e^{i\gamma_{\mu\nu}}+\hc+(\mu\rightleftharpoons\nu)\right)
+J_{\pm\pm}\left(S_{\vec{x}\mu}^{+}S_{\vec{x}\nu}^{+}e^{-i\gamma_{\mu\nu}}+S_{\vec{x}\mu}^{+}S_{\vec{x}+\vec{\mu}-\vec{\nu},\nu}^{+}e^{-i\gamma_{\mu\nu}}+\hc\right) \right].
\end{aligned}
\end{eqnarray}
\end{widetext}
This gives the full non-resonant Raman vertex, $\mathcal{R}$, for incoming light
with polarization $\vhat{e}_i$ scattered to light
with polarization $\vhat{e}_f$ as
\begin{equation}
\label{ramanvertexfinal}
\mathcal{R} \equiv \sum_{\mu<\nu}[(\vhat{e}_{i}\cdot\vec{\mu})(\vhat{e}_{f}\cdot\vec{\mu})+(\vhat{e}_{i}\cdot\vec{\nu})(\vhat{e}_{f}\cdot\vec{\nu})]{\mathcal{R}}_{\mu\nu}.
\end{equation}
Using the definition of the lattice vectors $\vec{\mu}$
(see Appendix \ref{app:defn}), the Raman vertex can be re-written as
\begin{equation}
  \label{ramanvertexchannels}
  \mathcal{R} =
  \frac{1}{8}\vhat{e}_i \cdot \left(\begin{array}{ccc}
                           \sum_{\mu<\nu} \mathcal{R}_{\mu\nu} 
                        & \mathcal{R}_{03}-\mathcal{R}_{12} &  \mathcal{R}_{02}-\mathcal{R}_{13} \\
           \mathcal{R}_{03}-\mathcal{R}_{12}  & \sum_{\mu<\nu} \mathcal{R}_{\mu\nu}  &  \mathcal{R}_{01}-\mathcal{R}_{23} \\
           \mathcal{R}_{02}-\mathcal{R}_{13} & \mathcal{R}_{01}-\mathcal{R}_{23} & \sum_{\mu<\nu} \mathcal{R}_{\mu\nu} 
  \end{array}\right) \cdot \vhat{e}_f.
\end{equation}
We see that the vertex naturally breaks into two channels corresponding to the irreducible representations $A_{1g}$ and $T_{2g}$ of the  point group $O_{h}$ of the pyrochlore lattice. The fully symmetric
$A_{1g}$ operator is given by
\begin{align}
  {\mathcal{R}}_{A_{1g}}&\equiv \sum_{\mu<\nu} \mathcal{R}_{\mu\nu}  ={\mathcal{R}}_{01}+{\mathcal{R}}_{02}+{\mathcal{R}}_{03}+{\mathcal{R}}_{12}+{\mathcal{R}}_{13}+{\mathcal{R}}_{23},
\end{align}
while the three components of the $T_{2g}$ channels are given by
\begin{subequations}
\begin{align}
{\mathcal{R}}_{T^x_{2g}}&\equiv{\mathcal{R}}_{01}-{\mathcal{R}}_{23}, \\
 {\mathcal{R}}_{T^y_{2g}}&\equiv{\mathcal{R}}_{02}-{\mathcal{R}}_{13},\\
 {\mathcal{R}}_{T^z_{2g}}&\equiv{\mathcal{R}}_{03}-{\mathcal{R}}_{12}.
\end{align}  
\end{subequations}
At this level of approximation, the Raman vertex in the $A_{1g}$
channel is proportional to the Hamiltonian (\ref{H_0}) and, thus, does
not induce any inelastic transitions. Therefore, in what follows, we
focus on the $T_{2g}$ channel, which is the only active Raman channel
in the QSI within the approach developed here (see Sec.~\ref{sec:disc:micro} for
a discussion of some limitations).  More compactly, dropping
the $\mathcal{R}_{A_{1g}}$ operator, the effective Raman operator
$\mathcal{R}$ is then
\begin{equation}
  \label{ramanvertext2g}
  \mathcal{R} \sim \sum_{\alpha\beta\gamma} |\epsilon_{\alpha\beta\gamma}|
  \left(\hat{e}^\alpha_i \hat{e}^\beta_f+\hat{e}^\beta_i \hat{e}^\alpha_f\right) \mathcal{R}_{T^\gamma_{2g}}.
\end{equation}
We next outline how to compute the Raman response in the $T_{2g}$ channel. 

\begin{figure}[tp]\label{Feynmandiagrams}
  \includegraphics[width=0.9\columnwidth]{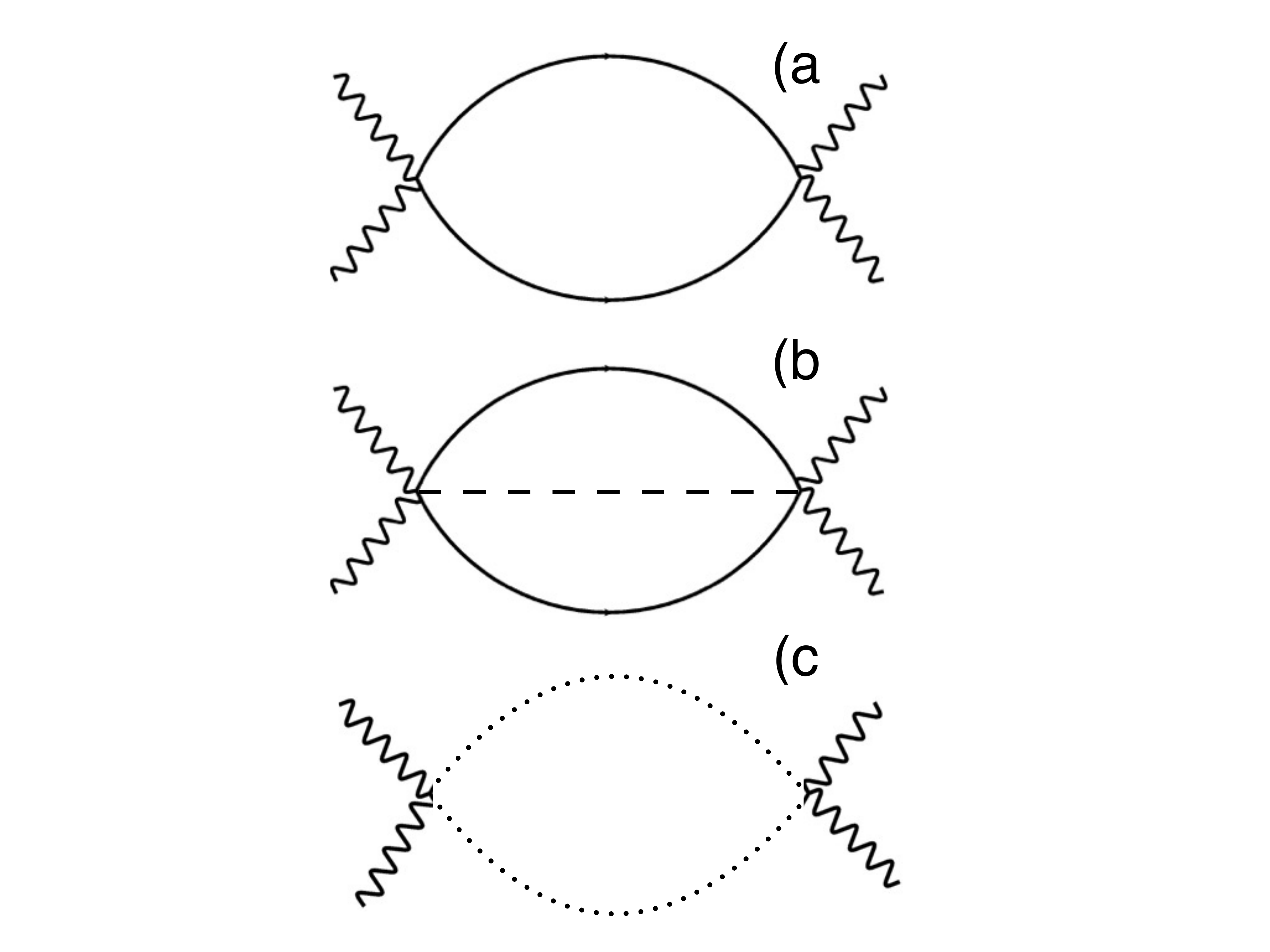}
\caption{Three classes of diagrams needed for calculating the Raman
scattering intensity from various QSI excitations: (a) spinon
excitations, (b) coupled spinon and gauge field excitations, (c)
$e$-excitations.  Solid, dashed and dotted lines represent
$G_{\sS}(\omega,\vec{k})$, $G_{A}(\omega,\vec{p})$ and
$G_{E}(\omega,\vec{p})$ propagators for the spinons, the $a$- and the
$e$-fields, respectively.  The wavy lines correspond to the incoming
and outgoing photons.
}
\end{figure}

\section{Raman response in Quantum spin ice}
\label{sec:raman-qsi}
We now compute the Raman scattering intensity at energy
transfer $\Omega \equiv \omega_i-\omega_f$ using Fermi's
golden rule, which is proportional to
\begin{equation}
  \label{Ramanintensity}
  \mathcal{I}(\Omega) \equiv \sum_{n} |\braket{n|\mathcal{R}|0}|^2 \delta\left(\Omega - E_n+E_0\right),
\end{equation}
where $\mathcal{R}$ is the Raman operator given in Eq.~(\ref{ramanvertext2g})
and $E_n$, $\ket{n}$ are the energies and eigenstates of the system.
Also, since we  are interested  the Raman response at zero-temperature we only
have intensity for positive $\Omega$. It is convenient to compute the Raman response using the time-ordered Raman response correlation function
\begin{equation}
  F(\Omega) \equiv \int dt\ e^{i\Omega t}\left[-i
    \avg{\mathcal{T} \mathcal{R}(t)\, \mathcal{R}(0)}\right],
\end{equation}
where $\avg{\cdots}$ is the average is with respect to the ground state.
In the spectral representation   $F(\Omega)$ can be written as
\begin{equation}
  F(\Omega) \equiv \sum_n \left[
    \frac{ |\braket{n|\mathcal{R}|0}|^2 }{\Omega - E_n+E_0 + i\delta} -
    \frac{ |\braket{n|\mathcal{R}|0}|^2 }{\Omega +E_n-E_0 -i\delta}
  \right],
\end{equation}
where $\delta = 0^+$. We then simply have that the intensity is given by
\begin{equation}
  \mathcal{I}(\Omega) = \frac{1}{\pi}\Theta(\Omega)\, \im  F(\Omega), 
\end{equation}
where $\Theta(\Omega)$ is the Heaviside function. It is convenient to
define a more generalized response tensor
\begin{eqnarray}\label{D}
  \mathcal{D}_{\mu\nu,\mu'\nu'}(t)=
  -i\avg{ \mathcal{T}{\mathcal{R}}_{\mu\nu}(t){\mathcal{R}}_{\mu'\nu'}(0)}
\end{eqnarray}
from which we can assemble the physical intensity of interest. Similarly
we can define a generalized intensity
\begin{equation}\label{Ramanintensity1}
\begin{aligned}
{\mathcal I}_{\mu\nu,\mu'\nu'} (\Omega)
%&\equiv \frac{1}{\pi}\Theta(\Omega) \im \left[ \int dt\ e^{i\Omega t}\
%\avg{\mathcal{T}{\mathcal{R}}_{\mu\nu}(t){ \mathcal{R}}_{\mu'\nu'}(0)}\right]\\&
=\frac{1}{\pi}\Theta(\Omega) \im \left[\int dt\ e^{i\Omega t}  \mathcal{D}_{\mu\nu,\mu'\nu'}(t)\right].
\end{aligned}
\end{equation}
For example, in terms of this generalized intensity the response in
the $T^x_{2g}$ channel is given by
\begin{equation}
  \label{eq:t-channel}
  \mathcal{I}_{T^x_{2g}}(\Omega) \equiv
  {\mathcal I}_{01,01} (\Omega)-
  {\mathcal I}_{01,23} (\Omega)-
  {\mathcal I}_{23,01} (\Omega)+
  {\mathcal I}_{23,23} (\Omega) . 
\end{equation}
Since the QSI state  is fully symmetric, the intensity
in each of the $T^\alpha_{2g}$ channels will be identical. We thus use
a common notation $\mathcal{I}_{T^x_{2g}}=\mathcal{I}_{T^y_{2g}}=\mathcal{I}_{T^z_{2g}}\equiv
\mathcal{I}_{T_{2g}}$.

To aid in the interpretation of these results, we divide the Raman
operators and the intensities into several parts that represent
qualitatively distinct physical processes. As for the $H_{\rm QSI}$,
Eq.~(\ref{HXXZ}) Hamiltonian, we can write the Raman operator in terms
of the spinon and gauge-fields.  We thus write the generalized Raman
operator, $\mathcal{R}_{\mu\nu}$, as a sum of several contributions
\begin{align}
  \mathcal{R}_{\mu\nu} &\equiv
                         \mathcal{R}_{\mu\nu}^{\sS}+
                         \mathcal{R}_{\mu\nu}^{\sSA}+
                         \mathcal{R}_{\mu\nu}^{\sE}.
\end{align}
There are three parts here: the scattering from spinons,
$\mathcal{R}^{\sS}$, the scattering from a combination of spinon and
gauge excitations, $\mathcal{R}^{\sSA}$, and the scattering from the
emergent light itself, $\mathcal{R}^{\sE}$.  In computing the Raman
operator, we consider only the leading terms of the exponential of the
gauge field, as we did in Eq.~(\ref{HXXZ}). These parts are closely
related to the decomposition of $H_{\rm QSI}$ [Eq.~(\ref{HXXZ})] into $H_{\sS}$ and
$H_{\sSA}$.

We begin with the terms of the Raman operator in
Eq.~(\ref{Roperator}) proportional to $J_{\pm}$ which can be written
as spinon-only and spinon-gauge interactions.  These terms can be expressed
analogously to  $H_{\sS}$ and $H_{\sSA}$ [see
Eqs.~(\ref{eq:raman-s}) and (\ref{eq:raman-sa})] as
\begin{subequations}  
\begin{align}
  \mathcal{R}^{\sS}_{\mu\nu} &= -\tilde{j}_{\pm}\sum_{\vec{k}\lambda} f^{\sS}_{\mu\nu}({\vec{k}})
                 \h{\psi}_{\vec{k}\lambda} \psi^{}_{\vec{k}\lambda}, \\
  \mathcal{R}^{\sSA}_{\mu\nu}&= -\frac{\tilde{j}_{\pm}}{\sqrt{N}}
  \sum_{\vec{k}\vec{p}\lambda}\sum_{\rho}
  f^{\sSA}_{\mu\nu\rho,\lambda}(\vec{k},\vec{p}) 
                             \h{\psi}_{\vec{k}+\vec{p},\lambda}\psi^{}_{\vec{k}\lambda}   A^{}_{\vec{p},\rho},
\end{align}
where the vertex $f^{\sS}_{\mu\nu}(\vec{k})$ is defined in Eq.~(\ref{eq:form-s}) and the
vertex $f^{\sSA}_{\mu\nu\rho,\lambda}(\vec{k},\vec{p})$ is defined in Eq.~(\ref{eq:form-sa}).

Next we consider the terms of the Raman operator [Eq.~(\ref{Roperator})] that are proportional to $J_{zz}$.
These terms  can be written using the electric field operators $E_{\vec{x}\mu}$ as
\begin{align}
  \mathcal{R}^{\sE}_{\mu\nu}
&\equiv  \sum_{\vec{x}\in \avg{A}} \left[
    E_{\vec{x}\mu}E_{\vec{x}\nu}+
    E_{\vec{x}\mu}E_{\vec{x}+\vec{\mu}-\vec{\nu},\nu}    
    \right],  \nonumber \\
  &= \sum_{\vec{k}} f^{\mu\nu}_E(\vec{k}) E_{\vec{k},\mu}E_{-\vec{k},\nu},
  \label{eq:raman-e}  
\end{align}
\end{subequations}
where we have defined the vertex
\begin{align}
    \label{eq:form-e}
  f^{\sE}_{\mu\nu}(\vec{k}) \equiv 1 + e^{-i\vec{k}\cdot(\vec{\mu}-\vec{\nu})}.
\end{align}
Note that, in contrast to the $J_{zz}$ parts of the Hamiltonian, since
the Raman operator contains $\sim \sum_{\vec{x}}S^z_{\vec{x}\mu}
S^z_{\vec{x}\nu}$, not summed over $\mu,\nu$, one cannot easily
represent this operator in terms of the charges $Q_{\vec{x}}$.

At leading order, these three parts do not mix since they create distinct sets of excitations.
We now calculate the intensity for each of these different mechanisms.
\subsection{Spinon-only  contribution}

This process involves the light scattering from a pair of
spinons. Since the incoming light carries (essentially) zero momentum,
this pair of particles must also have zero total momentum. The
spinon-only part of the response tensor is given by
\begin{align}
  \mathcal{D}^{\sS}_{\mu\nu,\mu'\nu'}(t)
  &\equiv
    -i\avg{\mathcal{T} \mathcal{R}^{\sS}_{\mu\nu}(t)\mathcal{R}^{\sS}_{\mu'\nu'}},\\
  &= -i \tilde{j}_{\pm}^2
  \sum_{\vec{k}\lambda,\vec{k}'\lambda'}
  f_{\mu\nu}^{\sS}(\vec{k}) f_{\mu'\nu'}^{\sS}(\vec{k}'), \nonumber \\
&\times \avg{\mathcal{T} \h{\psi}_{\vec{k}\lambda}(t) \psi^{}_{\vec{k}\lambda}(t)    \h{\psi}_{\vec{k'}\lambda'} \psi^{}_{\vec{k'}\lambda'}},
\end{align}
where $\mathcal{R}^{\sS}_{\mu\nu}$ is defined in Eq.~(\ref{eq:raman-s})
and $f^{\sS}_{\mu\nu}(\vec{k})$ in Eq.~(\ref{eq:form-s}).
%We can ignore the disconnected part of the operator; since the Raman
%vertex transforms as $T_{2g}$ it will not have a finite expectation
%value in a symmetry unbroken state like QSI. 
Using Eq.~(\ref{eq:green-s}), we can write the response tensor as
\begin{align}
  \mathcal{D}^{S}_{\mu\nu,\mu'\nu'}(t)
  &=
    2 i \tilde{j}_{\pm}^2  
    \sum_{\vec{k}}   f_{\mu\nu}^{\sS}(\vec{k}) f_{\mu'\nu'}^{\sS}(\vec{k})
    {G_{\sS}(\vec{k},t)} G_{\sS}(\vec{k},t),
\end{align}
where $G_{\sS}(\vec{k},t)$ is the Green's function of the spinon in
real-time.  The product of vertices
$f^{\sS}_{\mu\nu}(\vec{k})$ can be simplified to
\begin{equation}
      f_{\mu\nu}^{\sS}(\vec{k}) f_{\mu'\nu'}^{\sS}(\vec{k})=
  4
  \cos\left[\vec{k}\cdot(\vec{\mu}-\vec{\nu})\right]
  \cos\left[\vec{k}\cdot(\vec{\mu}'-\vec{\nu}')\right].
\end{equation}
The intensity tensor for the spinon-only  contribution is thus given
by
\begin{align}
  \mathcal{I}_{\mu\nu,\mu'\nu'}^{\sS}(\Omega)
  &\equiv \frac{1}{\pi} \Theta(\Omega) \,\im
    \int dt\ e^{i \Omega t} \mathcal{D}_{\mu\nu,\mu'\nu'}^S(t), \nonumber \\
  &=
    2 \tilde{j}^2_{\pm}
    \sum_{\vec{k}} \frac{
    f_{\mu\nu}^{\sS}(\vec{k}) f_{\mu'\nu'}^{\sS}(\vec{k})}
    {4 E_{\vec{k}}^2}
    \delta(\Omega-2E_{\vec{k}}),
\end{align}
where the Green's function integrals are
computed using conventional techniques.
To obtain the response in the $T_{2g}$ channel
we  use  Eq.~(\ref{eq:t-channel})  and get
\begin{align}
  \mathcal{I}_{T_{2g}}^{\sS}(\Omega)
  &=   \tilde{j}^2_{\pm} 
    \sum_{\vec{k}} \frac{8}{E_{\vec{k}}^2} \sin^2\left(\frac{k_y}{2}\right)\sin^2\left(\frac{k_z}{2}\right)
    \delta(\Omega-2E_{\vec{k}}) \\
  &=   \tilde{j}^2_{\pm} 
    \sum_{\vec{k}} \frac{8}{E_{\vec{k}}^2} \left[\frac{1}{3}\sum_{\alpha <\beta}\sin^2\left(\frac{k_\alpha}{2}\right)\sin^2\left(\frac{k_\beta}{2}\right)\right]
    \delta(\Omega-2E_{\vec{k}}).  \nonumber
\end{align}
In the last step we have symmetrized the vertex to emphasize
that this intensity is valid for each of the $R_{T^\alpha_{2g}}$ Raman
operators.
% Aside from the vertices, one can also see clearly
%that this is probing the zero total momentum spinon pairs.

\subsection{Spinon-gauge contribution}
Next we consider the spinon-gauge contribution.
Here, in addition to exciting a spinon-pair, the light also
excites an emergent photon. Due to the accompanying emergent
photon, the spinon-pair can have arbitrary total momentum.
To evaluate the intensity, we consider the
response tensor
\begin{equation}
  \mathcal{D}^{\sSA}_{\mu\nu,\mu'\nu'}(t)=-i\avg{\mathcal{T} \mathcal{R}^{\sSA}_{\mu\nu}(t) \mathcal{R}^{\sSA}_{\mu'\nu'}(0)},
\end{equation}
where $\mathcal{R}^{\sSA}_{\mu\nu}$ is defined in Eq.~(\ref{eq:raman-sa}).
 Next step in computing  $\mathcal{D}^{\sSA}_{\mu\nu,\mu'\nu'}(t)$
 is the  evaluation of 
\begin{equation*}
  -i\avg{\mathcal{T}
    \h{\psi}_{\vec{k}+\vec{p},\lambda}(t)\psi^{}_{\vec{k}\lambda}(t)  A^{}_{\vec{p},\rho}(t)
    \h{\psi}_{\vec{k}'+\vec{p}',\lambda'}\psi^{}_{\vec{k}'\lambda'}   A^{}_{\vec{p}',\rho'}   }.
\end{equation*}
There is one relevant contraction, which yields
\begin{equation}
  - \delta_{\vec{p},-\vec{p}'} \delta_{\vec{k}',\vec{k}+\vec{p}}
\delta_{\vec{k},\vec{k}'+\vec{p}'}
\delta_{\lambda\lambda'}\delta_{\rho\rho'} G_{\sS}(\vec{k},t) G_{\sS}(\vec{k}+\vec{p},t) G_A(\vec{p},t),
\end{equation}
where $G_{\sS}(\vec{k},t)$ and $G_A(\vec{p},t)$ are the
Green's function of the spinon and gauge-field, respectively, as
defined in Eqs.~(\ref{eq:green-s}) and~(\ref{eq:green-a}).
The response tensor $  \mathcal{D}^{\sSA}_{\mu\nu,\mu'\nu'}(t)$ is then
given by
\begin{equation*}
  -\frac{\tilde{j}_{\pm}^2}{N} \sum_{\vec{k}\vec{p}}
  \Phi_{\mu\nu,\mu'\nu'}(\vec{k},\vec{p})
 G_{\sS}(\vec{k},t) G_{\sS}(\vec{k}+\vec{p},t) G_A(\vec{p},t),
\end{equation*}
where we have defined the vertex
\begin{equation*}
  \Phi_{\mu\nu,\mu'\nu'}(\vec{k},\vec{p})=    \sum_{\lambda\rho}
    \cc{\left[f^{\sSA}_{\mu\nu\rho,\lambda}(\vec{k},\vec{p})\right]}
    f^{\sSA}_{\mu'\nu'\rho,\lambda}(\vec{k},\vec{p}).
\end{equation*}
Performing the time integral [see Eq.~(\ref{Ramanintensity1})] using standard contour
methods, we obtain the  intensity  tensor in the  spinon and gauge-field channel to be
\begin{align}
  \mathcal{I}^{\sSA}_{\mu\nu,\mu'\nu'}(\Omega) =
  \frac{\tilde{j}_{\pm}^2}{N}\sum_{\vec{k}\vec{p}}
  & \Phi_{\mu\nu,\mu'\nu'}(\vec{k},\vec{p})
    \frac{U}{8 E_{\vec{k}} E_{\vec{k}+\vec{p}} \epsilon_{\vec{p}}} \nonumber \\
  &\times \delta(\Omega - (E_{\vec{k}} +E_{\vec{k}+\vec{p}}+\epsilon_{\vec{p}})).
\end{align}
For the $T_{2g}$ channel, the
vertex is given by
\begin{equation}
\Phi_{T_{2g}}(\vec{k},\vec{p}) = 16\left(1 - \cos{\left[k_y +\frac{p_y}{2}\right]}\cos{\left[k_z + \frac{p_z}{2}\right]}\right).
\end{equation}
The final result for the spinon-gauge contribution
to the $T_{2g}$ intensity is then
\begin{align}
  \mathcal{I}^{\sSA}_{T_{2g}}(\Omega) \nonumber
  = \frac{\tilde{j}_{\pm}^2}{N}\sum_{\vec{k}\vec{p}}
   & \frac{2 U}{E_{\vec{k}}E_{\vec{k}+\vec{p}}\epsilon_{\vec{p}}}
     \delta(\Omega-(E_{\vec{k}} +E_{\vec{k}+\vec{p}} + \epsilon_{\vec{p}}))\\
   &\times
     \left(1 - \cos{\left[k_y +\frac{p_y}{2}\right]}\cos{\left[k_z + \frac{p_z}{2}\right]}\right).     
     \nonumber  
\end{align}
We see that if the photon energy scale is small, the intensity is
proportional to the density of states of the spinon pairs with
arbitrary total momentum. As in the spinon-only case, we rewrite the
intensity $\mathcal{I}^{\sSA}_{T_{2g}}(\Omega)$ in a manifestly
symmetric form as
\begin{align}
  \mathcal{I}^{\sSA}_{T_{2g}}(\Omega) \nonumber
  = \frac{\tilde{j}_{\pm}^2}{N}\sum_{\vec{k}\vec{p}}
   & \frac{2 U}{E_{\vec{k}}E_{\vec{k}+\vec{p}}\epsilon_{\vec{p}}}
       \delta(\Omega-(E_{\vec{k}} +E_{\vec{k}+\vec{p}} + \epsilon_{\vec{p}}))\\
   &\times
\left(1 - \frac{1}{3}\sum_{\alpha<\beta}\cos{\left[k_\alpha +\frac{p_\alpha}{2}\right]}\cos{\left[k_\beta + \frac{p_\beta}{2}\right]}\right).\nonumber  
\end{align}

\subsection{Electric field contribution}

Finally, we consider the contribution from the electric field alone.
Physically, this process corresponds to the light exciting a pair
of emergent photons. As in the case of spinons alone, the pair of emergent photons has
zero total momentum. The relevant response tensor is
\begin{align}
  \mathcal{D}^E_{\mu\nu,\mu'\nu'}(t)
  &\equiv
    -i\avg{\mathcal{T}\mathcal{R}^E_{\mu\nu}(t) \mathcal{R}^E_{\mu'\nu'}} \\
  &=
    \sum_{\vec{p}\vec{p}'} f_E^{\mu\nu}(\vec{p})f_E^{\mu'\nu'}(\vec{p}')
    \left[-i\avg{\mathcal{T}
    E_{\vec{p}\mu}(t)E_{-\vec{p}\nu}(t)
    E_{\vec{p}'\mu'}E_{-\vec{p}'\nu'}}\right] ,\nonumber
\end{align}
where $\mathcal{R}^E_{\mu\nu}$ is defined in Eq.~(\ref{eq:raman-e}).
This correlation function has two relevant contractions
leading to
\begin{align}
  \mathcal{D}^E_{\mu\nu,\mu'\nu'}(t)
  &=i
    \sum_{\vec{p}}
    \left[
    \delta_{\mu\nu'} \delta_{\mu'\nu} +\delta_{\mu\mu'} \delta_{\nu\nu'} \right]
    \left|f_E^{\mu\nu}(\vec{p})\right|^2
    G_E(\vec{p},t)^2, \nonumber
\end{align}
where the Green's function for the electric field, $G_E(\vec{k},t)$, is defined
in Eq.~(\ref{eq:green-e}). 
Evaluating the time integral  [see Eq.~(\ref{Ramanintensity1})] one finds
\begin{equation}
  \frac{1}{\pi}\im \int dt\ e^{i\Omega t} i G_E(\vec{p},t)^2 =
  \frac{\epsilon_{\vec{p}}^2}{4U^2}
  \left[
    \delta(\Omega+2\epsilon_{\vec{p}})
    +\delta(\Omega-2\epsilon_{\vec{p}})    
  \right]. \nonumber
\end{equation}
We thus have the generalized intensity from the
electric fields 
\begin{align}
\mathcal{I}^{E}_{\mu\nu,\mu'\nu'}(\Omega)=
  \sum_{\vec{p}}
  &
    \frac{\epsilon_{\vec{p}}^2}{4U^2}                  
    \left[
    \delta_{\mu\nu'} \delta_{\mu'\nu} +\delta_{\mu\mu'} \delta_{\nu\nu'} \right]
    \left|f_E^{\mu\nu}(\vec{p})\right|^2 \nonumber 
    \delta(\Omega-2\epsilon_{\vec{p}}).
\end{align}
The intensity can easily be evaluated for the $T_{2g}$ channel, yielding
\begin{align}
\mathcal{I}^{E}_{T_{2g}} (\Omega)&=
\sum_{\vec{p}}
\frac{\epsilon_{\vec{p}}^2}{U^2} 
\left[1+\cos\left(\frac{p_y}{2}\right)\cos\left(\frac{p_z}{2}\right)\right]
                \delta(\Omega-2\epsilon_{\vec{p}})\\
&=\sum_{\vec{p}}
\frac{\epsilon_{\vec{p}}^2}{U^2} 
\left[1+\frac{1}{3}\sum_{\alpha<\beta}\cos\left(\frac{p_\alpha}{2}\right)\cos\left(\frac{p_\beta}{2}\right)\right]
    \delta(\Omega-2\epsilon_{\vec{p}}) . \nonumber
\end{align}
This intensity reflects the density of states of a pair of emergent
photons with total momentum zero.  As in the spinon-only and
spinon-gauge cases, we have given the symmetric form for this
intensity.
\begin{widetext}

\subsection{Total intensity}
The total intensity in the $T_{2g}$ channel is thus given by
the following sum:
\begin{equation}
  \mathcal{I}^{}_{T_{2g}}(\Omega)=
  \mathcal{I}^{\sS}_{T_{2g}}(\Omega)+
  \mathcal{I}^{\sSA}_{T_{2g}}(\Omega)+
  \mathcal{I}^{\sE}_{T_{2g}}(\Omega),
\end{equation}
where the three different contributions are given by
\begin{subequations}
  \label{eq:intensities}
  \begin{align}
    \label{eq:intensity-s}
    \mathcal{I}^{\sS}_{T_{2g}}(\Omega)
  &=
    \tilde{j}_{\pm}^2 \sum_{\vec{k}}\frac{8}
    {E_{\vec{k}}^2}\left[\frac{1}{3}
    \sum_{\alpha<\beta}\sin^2\left(\frac{k_\alpha}{2}\right)\sin^2\left(\frac{k_\beta}{2}\right)\right]
    \delta(\Omega-2E_{\vec{k}}),\\
    \label{eq:intensity-sa}
  \mathcal{I}^{\sSA}_{T_{2g}}(\Omega) 
  &= { \tilde{j}_{\pm}^2}\sum_{\vec{k}}\left[\frac{1}{N}\sum_{\vec{p}}
    \frac{2 U }{E_{\vec{k}}E_{\vec{k}+\vec{p}}\epsilon_{\vec{p}}}
    \left(1 - \frac{1}{3}\sum_{\alpha<\beta}\cos{\left[k_\alpha +\frac{p_\alpha}{2}\right]}\cos{\left[k_\beta + \frac{p_\beta}{2}\right]}\right)   
    \delta(\Omega-(E_{\vec{k}} +E_{\vec{k}+\vec{p}} + \epsilon_{\vec{p}}))\right], \\
    \label{eq:intensity-e}
\mathcal{I}^{\sE}_{T_{2g}}(\Omega) &= 
\sum_{\vec{p}} 
\frac{\epsilon_{\vec{p}}^2}{U^2} 
\left[1+\frac{1}{3}\sum_{\alpha<\beta}\cos\left(\frac{p_\alpha}{2}\right)\cos\left(\frac{p_\beta}{2}\right)\right]
    \delta(\Omega-2\epsilon_{\vec{p}}).      
  \end{align}
\end{subequations}  
%Note that
%these intensities scale with the system size; this is due to the
%factor $g_i g_f$ omitted from the Raman operators. One has that $g_i
%g_f \sim 1/V$ where $V$ is the volume of the sample interacting with
%the light; including this would render the intensity explicitly
%intensive.
\end{widetext}
A brief comment is in order. We see that when ${\tilde j}_{\pm}=0$,
i.e. when the underlying system is simply a classical spin ice, the
contributions from the coupled spinon and gauge fields and from
spinons alone both vanish, ${\mathcal I}_{T_{2g}}^{\sSA}=0$ and
${\mathcal I}_{T_{2g}}^{\sS}=0$.  The intensity ${\mathcal
I}_{T_{2g}}^{E} $ from the electric part of the gauge field comes from
the $J_{zz}$ term and does not have explicit proportionality to
${\tilde j}_{\pm}$. Nevertheless, this contribution also vanishes when
${\tilde j}_{\pm}=0$; without the quantum tunneling terms, a spin
Hamiltonian containing only $J_{zz}$ terms commutes with the Raman
operator and, therefore, does not lead to a Raman response. More
explicitly, the photon dispersion collapses as ${\tilde j}_{\pm}
\rightarrow 0$, since $\epsilon_{\vec{k}} \propto {\tilde j}_{\pm}^3$,
leading to the zero response at finite frequency $\Omega$.

\begin{figure}
\includegraphics[width=0.99\columnwidth]{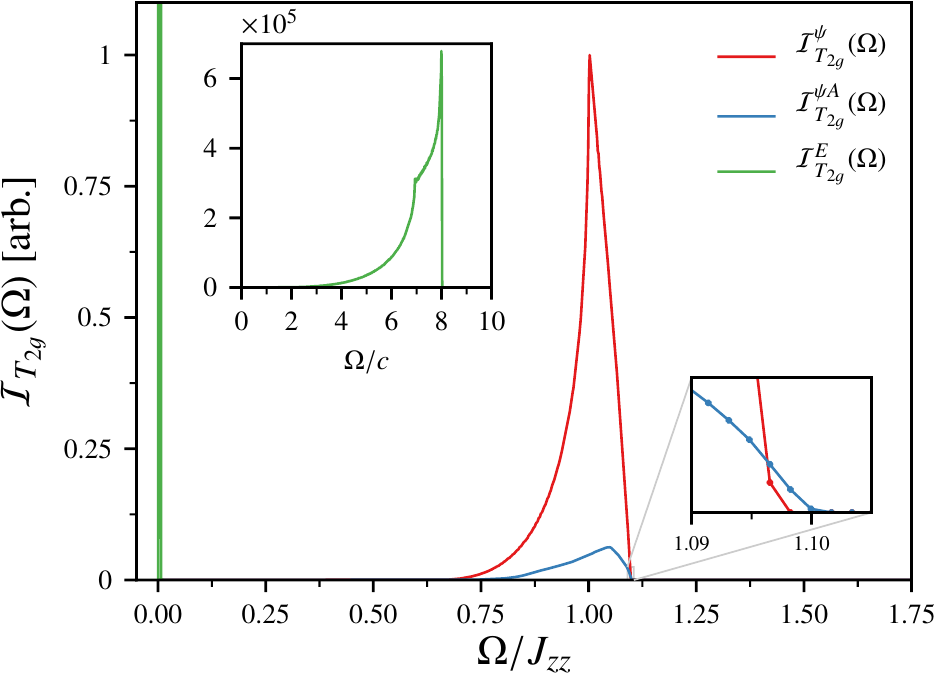}
\caption{The Raman intensities  in the  ${T_{2g}}$ polarization channel computed for QSI.
  Contributions from  the pure spinon contribution (blue line),  from the coupled spinon and gauge  fluctuations (red line) and   from  the gauge fluctuations coming from the $E$ field (green line)
  are shown, as defined in Eq.~(\ref{eq:intensities}). All intensities are normalized on the maximum intensity of the spinon-only Raman response, when taken alone.
  \label{Fig-Raman-intensity}
}
\end{figure}

\subsection{Numerical results}
\label{sec:results}
With  the developed formalism in hand, we now numerically evaluate the Raman
intensities in the $T_{2g}$ polarization channel. We examine the
contribution from each of the different physical processes: namely the
spinon-only, spinon-gauge and gauge-only contributions.  The single sums
found of Eqs.~(\ref{eq:intensity-s}) and (\ref{eq:intensity-sa}) were
evaluated on a grid of $384^3$ $\vec{k}$ points, with the origin shifted by a
small amount to resolve  any singularities in the
vertices at $\vec{k}=0$. For the
double sum of Eq.~(\ref{eq:intensity-e}), a similar procedure was
employed, but $48^3$ points for each momentum proved sufficient to
reach convergence.  The results for the Raman intensity profiles are
presented in Fig.~\ref{Fig-Raman-intensity}. This figure contains the main results
of this paper. All intensities are
computed assuming ${j}_{\pm}= 0.05$ (taking $\avg{s^{\pm}}=1$ for
simplicity) and are normalized to the maximum intensity of the
spinon-only Raman response.

\begin{figure*}[tp]
  \centering
  \includegraphics[width=0.31\textwidth]{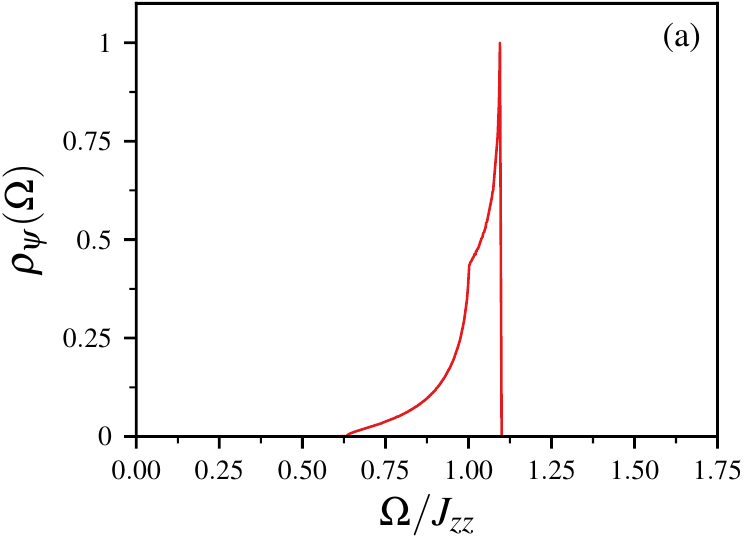}\hspace{0.5cm}
  \includegraphics[width=0.31\textwidth]{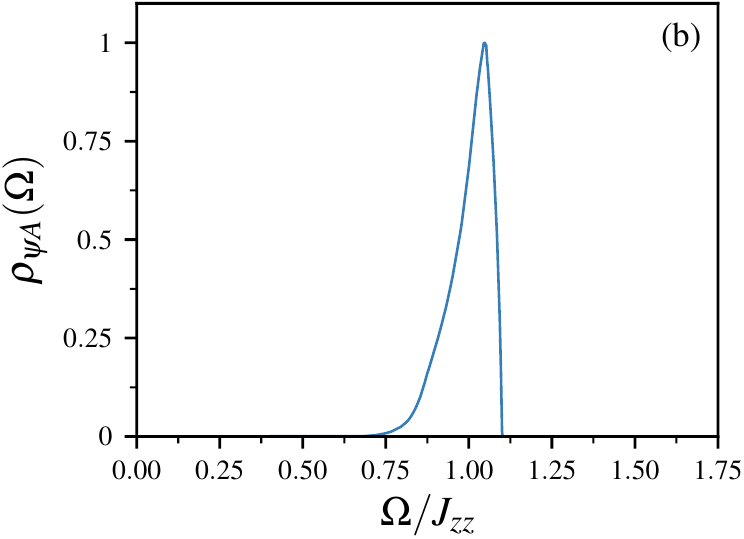}  \hspace{0.5cm}
  \includegraphics[width=0.31\textwidth]{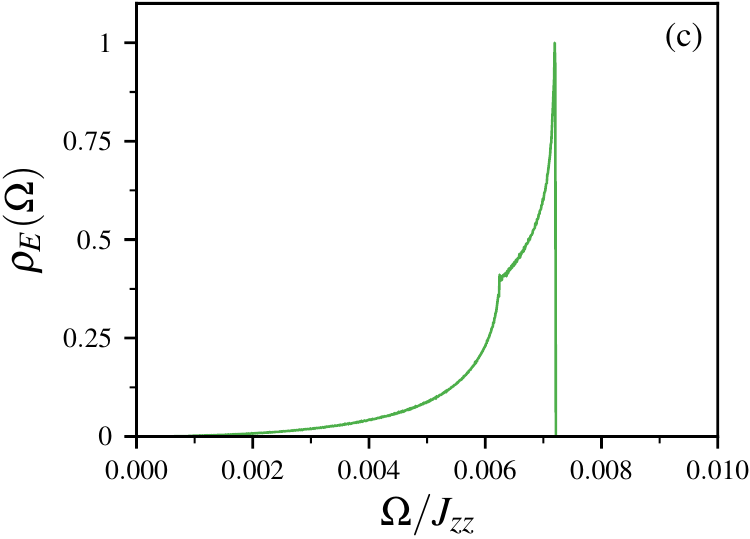}  
  \caption{
    \label{fig:dos}
    Density of states for (a) zero-momentum two-spinon states (see Eq.~(\ref{eq:spinon-dos})), (b) two-spinon
    states and  an emergent photon (see Eq.~( \ref{eq:spinon-gauge-dos})) and (c) two emergent photons alone (see Eq.~(\ref{eq:electric-dos})).
    Each density of states is normalized arbitrarily, such that its own individual maximum value is equal to one.
  }
\end{figure*}

First, we consider the Raman intensity from the spinon-only
scattering.  As expected, this contribution has intensity centered
around the classical spinon energy cost $\sim J_{zz}$ with a width
proportional to the energy of the tunneling term $J_{\pm}$.  Since the
incoming light can only generate a spinon pair with zero total
momentum, this channel does not probe the full two-spinon continuum.
Some aspects of the spinon-only response can be better understood by
considering the zero-momentum, two-spinon density of states defined as
\begin{equation}
  \label{eq:spinon-dos}
  \rho_\sS(\Omega) \propto \sum_{\vec{k}}\delta(\Omega-2E_{\vec{k}}),
\end{equation}
as shown in Fig.~\subref{fig:dos}{(a)}. Here we can see that the onset
of the spinon density of states is $\sim \sqrt{\Omega-\Omega_0}$ with
$\Omega_0 = \sqrt{1-12\tilde{j}_{\pm}}$ being twice the gap in the
spinon dispersion of Eq.~(\ref{spinondispersion}). Further, the
density of states also has a sharp peak (Van Hove singularity) due to
the presence of flat regions in the spinon dispersion.~\cite{Hao2014}
These features -- a slow onset at low frequencies and a large
intensity near the maximum of the two-spinon band -- are
characteristic features of the two-spinon Raman response.

The Raman response from coupled spinon and gauge fluctuations also
takes the form of a broad continuum in roughly the same range of
energies as the two-spinon case. However, the excitation of the
emergent gauge photon now allows access to the full two-spinon
continuum, as its presence relaxes the constraint on having zero total
momentum for the spinon pair.  Correspondingly, the Raman response in
this channel includes the full two-spinon continuum.  While this does
not change the maximum or minimum of the two-spinon energies (compared
to the zero-momentum case), it does affect the intensity profile at
intermediate energies. As in the spinon-only case the spinon-gauge intensity
can be better understood by considering the corresponding density of states
\begin{equation}
  \label{eq:spinon-gauge-dos}
  \rho_{\sSA}(\Omega) \propto \sum_{\vec{k} \vec{p}}\delta(\Omega-(E_{\vec{k}}+E_{\vec{k}+\vec{p}} + \epsilon_{\vec{p}})),
\end{equation}
as shown in Fig.~\subref{fig:dos}{(b)}. One important feature is that
the width of this broad continuum is slightly larger than that of the
pure two-spinon scattering; due to the interaction with the
gauge-field, the combined spinon-photon states can reach higher
energies than the spinons alone. This can be seen in the bottom right
inset of Fig.~\ref{Fig-Raman-intensity}, where the upper edge of the
intensity is pushed to higher energies. However, this shift is quite
small, being proportional to the emergent photon bandwidth, which
scales as $\sim j_{\pm}^3$.  Given this fact, we can effectively
ignore the energy of gauge particle in the $\delta$-functions in $
\mathcal{I}_{T_{2g}}^{\sSA}$. Physically, the photon is thus acting as
a ``momentum-sink'' for the spinon-pair: for essentially zero energy
cost one can excite a photon with arbitrary momentum.

Finally, we consider the gauge-field-only response from the emergent
electric field. It appears as a strong, sharp peak at the energies
corresponding to the emergent photon bandwidth.  The energy scale of
the emergent photon dispersion relation goes as $\sim j_{\pm}^3$ and
is thus much smaller than the energy scale of the aforementioned
features that involve the spinons. The intensity profile of the Raman
response in this case follows very closely the zero-momentum,
two-photon density of states
\begin{equation}
  \label{eq:electric-dos}
  \rho_E(\Omega) \propto \sum_{\vec{p}}\delta(\Omega-2\epsilon_{\vec{p}}),
\end{equation}
which is shown in Fig.~\subref{fig:dos}{(c)}. At low energies, this
intensity follows a power-law $\sim \Omega^2$ due to the linear photon
dispersion. The flat dispersion in the photon band structure at the
edge of the band (at $\Omega = 8c$) is also apparent high intensity at
the highest energies. The larger intensity relative to the spinon
features seen in Fig.~\ref{Fig-Raman-intensity} originates from the
lack of a $\sim j_{\pm}^2$ prefactor in the Raman intensity since this
scattering processes is due to the large $J_{zz}$ interactions, and
the narrow support in $\Omega$.

\section{Discussion}
\label{sec:discussion}
In this section we discuss some of the limitations of the results derived in
this work and how they may affect applications to real materials. In
particular, we discuss the microscopic origins of the Raman operator,
the approximations made in the slave-particle formulation and 
speculate on the effects of the so far ignored anisotropic $J_{\pm\pm}$ and
$J_{z\pm}$ interactions on the Raman intensity.

\subsection{Slave-particle formulation}
\label{sec:disc:slave}
In Secs. \ref{sec:qsi} and
\ref{sec:spinon-gauge} we introduced a slave-particle formulation for
QSI and a framework to enable calculation of the Raman response. There
are a number of approximations involved in the
slave-particle framework used in this work.

First, in our analysis, we considered only the first order coupling
between the spinons and gauge fluctuations, i.e. we kept only the
first order terms in the expansion of the exponential $e^{iA}$. This
approximation is not necessarily controlled. Indeed this approximation
removes completely the gauge monopoles,~\cite{Hermele2004} a set of
excitations of the gauge sector with energy comparable to that of the
emergent photon. These  excitations appear only  non-perturbatively in $A$. A detailed
analysis of corrections from the higher orders in the expansion of
$e^{iA}$ gauge sector and the possibility of including the gauge
monopoles are left for future exploration.

Second, we have computed the response for the XXZ model where only
$j_{\pm}$ is non-zero. The features seen in the intensities are likely
to be modified if the Raman response was computed for the complete and
more general anisotropic exchange model,
Eq.~(\ref{H_0})~\cite{Ross2011,Savary2012}.  The $J_{z\pm}$ terms
would bring another spinon-gauge field interaction vertex, and the
$J_{\pm\pm}$ terms would bring a four-spinon interaction vertex. A
detailed analysis of these vertices is, technically, significantly more
involved, and we therefore leave it for future study.  However,
believe that the basic qualitative features of the Raman response, the
broad intensity continuum and its width would not be changed
significantly by the inclusion of these interactions.~\footnote{Note
that a pure $J_{\pm\pm}$ coupling, with $J_{\pm}= J_{z\pm} = 0$ would
not produce spinon pairs, but two singly-charged spinons and one
doubly-charged spinon. Therefore in this case the intensity would
appear only near the much higher energy $3J_{zz}$.}
The detailed features, such as the sharpness of the peak in the
spinon-only response, would likely be modified.

Third, one key deficiency in the mean-field theory presented in
Sec.~\ref{sec:qsi} is that the effects of the gauge-field on the
spinon are treated in an averaged way. It is unclear whether such an
approximation is valid in QSI, given the energy scale of the photon is
much lower than the kinetic energy of the
spinons.~\cite{wan2016spinon} However, recent more exact treatments of
the spinon excitations in related
contexts,~\cite{wan2016spinon,kourtis2016,petrova2015} have found that
treating the spinon as a (strongly) renormalized free
particle,~\cite{wan2016spinon,kourtis2016} may not be too poor of an
approximation.

Finally, we note that the emergent photon-only response is derived entirely
from the gauge part of the model, $H_g$. This is essentially
equivalent to the lattice gauge theory of
Ref.~[\onlinecite{Benton2012}] used to describe the physics of QSI when
the spinons are not included. This description has proven to be
quite accurate in computing the static properties of QSI in this
limit, faithfully reproducing the results of direct
simulation..~\cite{Benton2012} We thus expect that our results for the low
energy, electric only part of the Raman intensity to be robust, as it
is independent of some of these coarser approximations used in the
slave-particle formulation.

\subsection{Microscopic considerations}
\label{sec:disc:micro}
In Sec.~\ref{sec:raman} we derived the Raman operator through
degenerate perturbation theory. In doing so we made several
approximations that simplify both the calculation and the form of the Raman
operator.

First, we comment on the generated polarization channels. Within the
approximations used, one obtains an inactive $A_{1g}$ channel and an
active $T_{2g}$ channel. The Raman operator, $\mathcal{R}$ [see
Eq.~(\ref{Roperator})] for each of these channels mimics closely the
``parent'' exchange model, $H_{\exch}$ [see Eq.~(\ref{H_0})], with the
different anisotropic interactions appearing in the same ratios as in
the exchange Hamiltonian. The appearance of these exchange constants
in the Raman operator is predicated on the assumption that one can
neglect the photon energies in the relevant denominators in the
perturbation theory. When this condition is relaxed, the form of the
Raman operator becomes decoupled from that of the exchange
Hamiltonian. Indeed, the rich structure of the intermediate states
involved in rare-earth super-exchange processes\cite{iwahara2016} will
likely affect not only the scale of the interactions, but also the
relative importance of the various anisotropic terms. Because of this,
we would expect additional polarization channels to be generated and
that the inactive $A_{1g}$ would become active.  Roughly, such
corrections would be proportional to $\omega_{i}/U_f$ where
$\omega_{i}$ is the incoming light frequency and $U_f$ is a typical
rare-earth charge-transfer energy scale.

Second, we note here that our treatment of the Raman intensity only
includes contributions from two-ion processes, whereas single-ion
processes, for example mediated through the $5d$ orbitals of the
rare-earth site or through the surrounding oxygen ions, have not been
included in our calculation. Since the energy cost of single-spin
flips and two-spin-flips can be of the same order in QSI materials,
these processes might also be important.  This can be contrasted with
the separation of energy scale in one- and two-magnon processes in
conventionally ordered magnets.~\cite{LF1968} For Kramers ions, any
single-ion Raman operator is necessarily time-reversal odd, and thus
must vanish as the frequency of the incoming light becomes small
relative to the atomic energy scales,~\cite{moriya1968theory}
providing some suppression of such contributions.  Further, the Raman
response will be non-zero only in the $T_{1g}$ channel, as
time-reversal odd operators only appear in anti-symmetric
channels~\cite{moriya1968theory}, which for $O_h$ there is only
$T_{1g}$. For non-Kramers ion, the single-ion transverse $S^{\pm}$
operators can appear in the Raman operator without such
suppression. Within the context of our calculations here, such
single-ion terms are most easily generated via second-order virtual
processes that only involve the surrounding oxygens atoms. For the two
axial oxygens we have included (see Fig.~\ref{spinicefigure} and
Sec.~\ref{sec:raman}), such a contribution vanishes (for details see
Appendix \ref{app:single-ion}). However, there are six additional,
lower symmetry oxygens that we have not considered.~\cite{Gardner2010}
If these are included, single-ion terms are generated and they
contribute to the Raman response in both the $E_g$ and $T_{2g}$
channels.  However, these may be somewhat suppressed given the larger
distance to these oxygens.~\cite{Gardner2010} Even given these
complications, one should note that these single-ion operators probe
the same excitations as the two-ion operators: the $S^{\pm}$ type
terms excite spinon-pairs, while the $S^z$ type terms excite emergent
photons.  We thus do not expect any qualitative change in the results
presented here for the $T_{2g}$ channel when such single-ion terms are
included.
\section{Conclusions}
\label{sec:conclusions}
In this paper, we proposed a theory of the Raman scattering in the XXZ
limit of the general anisotropic exchange model, which we analyzed using a slave-particle
formulation of QSI.\cite{Hao2014} We derived the Raman vertex
using the traditional framework of an effective Hamiltonian for the
interaction of light with spin degrees of
freedom.\cite{LF1968,Shastry1990,Shastry1991,Ko2010} We showed that,
at fourth order in perturbation theory, the Raman vertex of Eq.
(\ref{ramanvertexfinal}) takes a Loudon-Fleury form,~\cite{LF1968}
generated by photon-assisted super-exchange, following the anisotropic
exchange model [Eq.~(\ref{HXXZ})] that leads to the QSI behavior.  We
also showed that the Raman vertex naturally decomposes into two
channels corresponding to the irreducible representations $A_{1g}$ and
$T_{2g}$ of the lattice point group.  Moreover, since the Raman vertex
in the $A_{1g}$ channel commutes with the QSI Hamiltonian, the Raman
intensity is non-zero only in the $T_{2g}$ polarization channel.
Within this framework, we decomposed the Raman intensity into three
contributions, from the pure spinon field, coupled spinon and gauge
fluctuations and the emergent photon. We showed that the dominant
feature of the overall response consists of a broad continuum from the
two-spinon spectrum and a sharp narrow peak at low energy originating
from the gauge fluctuations of the emergent photon field taken alone.
To conclude, we comment below on a few aspects of Raman scattering
in general, as well as discuss relevance of these results to real
candidate QSI materials.

First, a unique feature of Raman scattering is the ability to probe
characteristics of the system that are not directly related to the
magnetic moments. For example, the QSI candidates to date have mostly
been studied with tools such as neutron scattering. \cite{Gingras2014}
While this approach has proven to be very powerful, there are some
limitations when the pseudo-spins are of dipolar-octupolar or
non-Kramers character, as they are in \dto{}, \hto{}, \tto{} and in
the \abo{Pr}{M} family.  For these compounds, the transverse
components of the pseudo-spin are higher multipoles (quadrupoles or
octupoles) and thus are not easily visible in neutron scattering. So
while inelastic neutron scattering could observe the photon excitation
in such materials (in principle, with sufficient energy
resolution),\cite{Benton2012} observing the spinon excitations is very
difficult.\footnote{
  For the case of dipolar-octupolar doublets,\cite{huang2014quantum} there are some possible
  routes around this restriction. In this case both $S^z$ and $S^x$ transform
  as dipoles oriented along the local $\vhat{z}$ direction and thus can mix.  
  In principle the exchange interactions could stabilize a spin ice
  in a mixed direction, i.e. $\sum_{\avg{ij}} \tilde{S}^z_i \tilde{S}^z_j$
  where $\tilde{S}^z_i = S^z_i \cos{\theta} +S^x_i \sin{\theta}$. In this
  case the moment operator, $\propto S^z$, then involves both $\tilde{S}^z$
  and $\tilde{S}^x$ and thus can probe the spinon excitations. The spinon excitations
  would also be visible in the so-called ``octupolar'' QSI introduced in Ref.~[\onlinecite{huang2014quantum}].
  In \dto{} the interactions between the dipoles is dominant
  and any mixing is thought to be minimal.\cite{Rau2015}
} The possibility of seeing the two-spinon
continuum at all, irrespective of resolving distinct signatures or
features, is from a fundamental perspective a strong asset for Raman scattering as a probe of QSI
candidate materials.

Second, from a broader perspective, we would like to comment on the
possibility to use Raman scattering as a tool to study the phase
transitions between different magnetic phases.  In particular, it
would be interesting to compare the Raman responses arising from a QSL
phase and nearby ordinary magnetically ordered phase appearing at
slightly different set of parameters of the same model.  For example,
aside from exotic phases such as QSI and the conjectured Coulomb
ferromagnet,\cite{Savary2012} there are also four magnetically
ordered phases found in the phase diagram of the anisotropic exchange
model. These are the antiferromagnetic $\Gamma_5$ states, a family of
splayed ferromagnets, the Palmer-Chalker state and the all-in/all-out
(AIAO) order~\cite{Hao2014}.  Even in the simple limit considered here with
$j_{\pm\pm}=j_{z\pm}=0$, there is the nearby $\Gamma_5$ state that is
stabilized for $j_{\pm} \gtrsim
0.06$.\cite{banerjee2008,kato2015numerical}  The transition into the
state can be captured with the slave-particle description used
here and  it corresponds to  the condensation of the
spinons,\cite{Savary2012} which is similar to the gauge symmetry breaking in the Higgs'
mechanism. More generally, near the boundary of a QSL and a
magnetically ordered phase, one may expect  both the
conventional excitations of the ordered phases and the unconventional
excitations of the QSL to be generically present. The ability to track
both the conventional and unconventional excitations across the phase
transition through their Raman response could prove useful in the
understanding of both phases and the transition itself.
%Whether the
%spinon-gauge-field formalism can still be productively applied within
%the ordered antiferromagnetic phase remains an open question, but if so,
%one may be able to compute the Raman response for the
%antiferromagnetic phase using the same description used for the QSI
%and then, the Raman response may be used as a tool to study the phase
%transition between these two phases.
%In general, in frustrated magnetic systems, both the response from the
%QSL and from the magnetically ordered state will have a broad Raman
%intensity background with some peak structure.
%\cite{Knolle2014,Brent2015,Brent2016-short,Brent2016-long,Perkins2008,Perkins2013}
%Thus, in order to understand the origin of the broad continua in the
%Raman spectrum, one needs to identify the nature of the excitations,
%disentangle the contributions of individual quasi-particles with
%different energies, and make a detailed, quantitative comparison
%between theory and experiment.  In all, this is not a trivial question
%to answer and detailed discussion on this issue is still lacking.

One last and yet very important question to address is the possibility
to see the Raman responses in experiments on real QSI materials. As
far as we know, no magnetic Raman experiments have been done on QSI
materials so far. One clear obstacle is the fact that the energy scale
of exchange interactions in the rare-earth magnets is considerably smaller
than  one in many transitional metal magnets. For the rare-earth pyrochlore
quantum spin ice materials, these coupling constants are typically on
the order of $0.1 {\rm meV}$.  Coupling constants of this magnitude
will produce the gross features (scattering from spinons) at energies
of order 1-2 cm$^{-1}$, which is, unfortunately, much smaller than the lower limit
accessible by current Raman spectroscopy, which typically probes
excitations ranging 1-100 meV (10-1000 cm$^{-1}$
).\cite{Devereaux2007} However, one possible way to resolve this
conundrum might be with Brillouin scattering, which does well for
probing energy scales 0.01-1 meV (0.1-10
cm$^{-1}$)\cite{Hayes2012,Polian2003} and which differs from Raman
scattering technique only by the type of spectrometer. However, even
with such a setup, the intensity due to scattering from the emergent
photon is likely to remain challenging to expose. Further
complications can arise in spin ice systems where the spin ice
manifold itself is split by dipolar interactions, such as in \dto{} or
\hto{}. This splitting carries over to the spinon (or classical
``monopole'') excitations and thus could mimic the effects of a
quantum dispersion in the Raman intensity.

On the other hand, as material science is a fast developing field of
research, we believe that new QSI materials with stronger quantum
effects may be designed or discovered. One tantalizing possibility
could be the discovery of a transition-metal quantum spin ice
candidate. If  such a system  were to exist, a large increase (one or two orders of magnitude) in
energy scale relative to the rare-earth materials considered in the present work could possibly
render many of the features discussed here at much more experimentally accessible
energies. In such a scenario, not only the spinon continuum, but the emergent photon itself
could even be visible within experimentally accessible energy ranges.

\section{Acknowledgments}
The  authors  are  grateful  to  Girsh Blumberg,  Kenneth Burch,  Zhihao Hao and Brent  Perreault for  helpful  discussions. N. P. and J. F. acknowledge the support from NSF
DMR-1511768 Grant.  N. P.  and J.F. are  thankful to the Perimeter Institute for hospitality during the
course  of  this  work.  Research  at  the
Perimeter  Institute  is  supported  by  the  Government  of
Canada through Industry Canada and by the Province
of  Ontario  through  the  Ministry  of  Economic  Development and Innovation. Part of this work was also performed at the Aspen Center for Physics, which is supported
by NSF Grant No. PHY-1066293. 
The work at the University of Waterloo was
supported by the Natural Sciences and Engineering Research
 Council of Canada (NSERC), the Canada Research Chair (CRC) program
(M.J.P.G., Tier 1) and the Canadian Institute for Advanced Research (CIfAR).

\appendix

\section{The definition of local coordinate space and the $\zeta$ matrix} \label{app:defn}
The definition of the lattice vectors $\vec{\mu}$ is: 
\begin{align}
\label{mu0123}
\vhat{0}&=\frac{+\vhat{x}+\vhat{y}+\vhat{z}}{4},&  
\vhat{1}&=\frac{+\vhat{x}-\vhat{y}-\vhat{z}}{4}, \nonumber \\
 \vhat{2}&=\frac{+\vhat{y}-\vhat{x}-\vhat{z}}{4}, &
 \vhat{3}&=\frac{+\vhat{z}-\vhat{x}-\vhat{y}}{4},
\end{align}
where $\vhat{x}, \vhat{y}, \vhat{z}$  denote the global cubic axes.
The local coordinates ($\vhat{x}_{\mu}$, $\vhat{y}_{\mu}$, $\vhat{z}_{\mu}$) for the four sites
(labeled as $\mu=0,1,2,3$) of a certain tetrahedron  of the pyrochlore lattice are defined as
\begin{subequations}
  \label{eq:local-basis}
\begin{align}
&\vhat{x}_{0}=\frac{-2\vhat{x}+\vhat{y}+\vhat{z}}{\sqrt{6}}, \, 
\vhat{y}_{0}=\frac{-\vhat{y}+\vhat{z}}{\sqrt{2}}, \, 
\vhat{z}_{0}=\frac{+\vhat{x}+\vhat{y}+\vhat{z}}{\sqrt{3}},\\
&\vhat{x}_{1}=\frac{-2\vhat{x}-\vhat{y}-\vhat{z}}{\sqrt{6}}, \,
\vhat{y}_{1}=\frac{+\vhat{y}-\vhat{z}}{\sqrt{2}}, \, 
\vhat{z}_{1}=\frac{+\vhat{x}-\vhat{y}-\vhat{z}}{\sqrt{3}},\\
&\vhat{x}_{2}=\frac{+2\vhat{x}+\vhat{y}-\vhat{z}}{\sqrt{6}}, \, 
\vhat{y}_{2}=\frac{-\vhat{y}-\vhat{z}}{\sqrt{2}}, \, 
\vhat{z}_{2}=\frac{-\vhat{x}+\vhat{y}-\vhat{z}}{\sqrt{3}},\\
&\vhat{x}_{3}=\frac{+2\vhat{x}-\vhat{y}+\vhat{z}}{\sqrt{6}}, \, 
\vhat{y}_{3}=\frac{+\vhat{y}+\vhat{z}}{\sqrt{2}}, \, 
\vhat{z}_{3}=\frac{-\vhat{x}-\vhat{y}+\vhat{z}}{\sqrt{3}}.
\end{align}
\end{subequations}

In Eq.~(\ref{H_0}), the phase factors $\gamma$ and $\zeta$ are defined as~\cite{Ross2011,Savary2012}
\begin{eqnarray}
\begin{aligned}
\gamma_{\mu\nu}=\left( \begin{array}{cccc} 0 & 1 & \omega & \omega^2 \\
1 & 0 & \omega^2 & \omega \\
\omega & \omega^2 & 0 & 1 \\
\omega^2 & \omega & 1 & 0 
\end{array} \right)_{\mu\nu},
\end{aligned}
\end{eqnarray}
where $\omega = e^{2\pi i/3}$ and $\zeta_{\mu\nu} = -\cc{\gamma}_{\mu\nu}$.
\section{Second order contributions to the Raman vertex}
\label{app:single-ion}
Here we consider the second order terms, $\sim PVRVP$, in the perturbative expansion
Eq. (\ref{perturbationexpansion}).  These processes can only result in
operators acting on a single rare-earth ion. Evaluating these terms within
the charging approximation~\cite{Onoda2011,Rau2015}, one finds
\begin{align}
{H}_{R}^{(2)}&={P}\VAn{1}{R}\VAn{1}{P}, \\
             &\sim -\left(\frac{e}{\hbar c}\right)^2\sum_{\alpha\beta\mu}(\vhat{e}_{i}\cdot\vec{\mu})(\vhat{e}_{f}\cdot\vec{\mu})
               \frac{2\left[t_{\mu} \h{t}_{\mu}\right]^{\alpha\beta}}{U_f}
          \sum_{\vec{x}\in\avg{A}}     P f^{}_{\vec{x}\mu,\beta} \h{f}_{\vec{x}\mu,\alpha}P,\nonumber
\end{align}
where, loosely, $U_f$ is an energy scale associated with the cost
transferring a hole from an oxygen to the rare-earth ion. As in
Sec.~\ref{sec:raman}, we including only hoppings to the high-symmetry
oxygens~\cite{Gardner2010} that sit at the centers of the rare-earth tetrahedra  (Wyckoff site $8b$). By
construction $H^{(2)}_R$ is symmetric in the polarizations
$\vhat{e}_i$ and $\vhat{e}_f$. Since coupling to time-reversal odd
operators must be in anti-symmetric channels,~\cite{moriya1968theory}
no time-reversal odd operators can be generated by this process. This
holds even when the charging approximation is lifted and when energy
of the light is included in the resolvents. For Kramers doublets, this
implies that $H^{(2)}_R$ does not contribute to the Raman
response. For non-Kramers doublets, this implies any operators
appearing $H^{(2)}_R$ must be time-reversal even. We can thus
(effectively) consider the operator
\begin{equation}
  P f^{}_{\vec{x}\mu,\beta} \h{f}_{\vec{x}\mu,\alpha}P \sim
  h^0_{\beta\alpha} + h^+_{\beta\alpha} S^-_{\vec{x}\mu} + h^-_{\beta\alpha} S^+_{\vec{x}\mu}
\end{equation}
For a given rare-earth site, the two high-symmetry oxygens are along the $\pm
\vhat{z}$ directions.  Because of this, within the Slater-Koster
(two-center) approximation~\cite{slater1954}, one then has that
$t^{}_{\mu} \h{t}_\mu$ is diagonal. The diagonal operators $P
f^{}_{\vec{x}\mu,\alpha} \h{f}_{\vec{x}\mu,\alpha}P$ are then
invariant under rotations about the $\vhat{z}_\mu$. Since the
$S^{\pm}_{\vec{x}\mu}$ are not invariant under these rotations, this
implies that $h^{\pm}_{\alpha\beta} = 0$. Note that this argument
holds even when the charging approximation is lifted and when energy
of the light is included, since the resolvents are invariant under
three-fold rotations about $\vhat{z}_\mu$.  We thus see that, within
our approximations, when only these two high symmetry oxygens~\cite{Gardner2010} are
included there is no second order, or single-ion, contributions to the
Raman response.

Note that this argument will not follow when the low-symmetry oxygens~\cite{Gardner2010}
(Wyckoff site $48f$) are included, and thus generically one has $h^{\pm}_{\alpha\beta} \neq
0$.  In addition, inclusion of additional hoppings in $t_{\mu}$,
beyond the Slater-Koster approximation, would render
$t^{}_{\mu}\h{t}_\mu$ non-diagonal and thus also give
$h^{\pm}_{\alpha\beta}\neq 0$.  We thus expect that for non-Kramers
ions one can have single-ion response from such operators.

\bibliography{QSI_refs}
\end{document}